%% file: main.tex
\documentclass{ article}

\usepackage{amsmath}
\usepackage{graphicx}
\usepackage{natbib}
\usepackage{subcaption}
\usepackage{comment}
\usepackage{enumitem}

\usepackage{PRIMEarxiv}

\usepackage[utf8]{inputenc} 
\usepackage[T1]{fontenc}    
\usepackage{hyperref}       
\usepackage{url}            
\usepackage{booktabs}       
\usepackage{amsfonts}       
\usepackage{nicefrac}       
\usepackage{microtype}      
\usepackage{lipsum}
\usepackage{fancyhdr}       
\usepackage{graphicx}       
\graphicspath{{media/}}     
\newtheorem{thm}{Theorem}
\newtheorem{theorem}{Theorem}

\input{macro.tex}

\def\bt{\begin{thm}}
\def\et{\end{thm}}

\def\be{\begin{eqnarray}}
\def\ee{\end{eqnarray}}
\def\bes{\begin{eqnarray*}}
\def\ees{\end{eqnarray*}}

\def\bfbeta{{\boldsymbol\beta}}

\def\parrow{\buildrel{ p}\over\longrightarrow}
\def\Darrow{\buildrel {\cal D}\over\longrightarrow}
\def\darrow{\buildrel d \over\longrightarrow}

\def\bfZ{{\bf Z}}
\def\bfz{{\bf z}}
\def\bfY{{\bf Y}}
\def\bfy{{\bf y}}
\def\Var{{\rm Var}}
\def\Cov{{\rm Cov}}

\def\cI{{\cal I}}

\def\bd{\begin{description}}
\def\ed{\end{description}}

\title{Covariate-adjusted Group Sequential Comparisons of Restricted Mean Survival Times
}

\author{
  Peter Zhang, Brent Logan, Michael Martens \\
  Division of Biostatistics \\
  Medical College of Wisconsin \\
  Milwaukee, Wisconsin, U.S.A
}

\begin{document}
\maketitle

\begin{abstract}
The restricted mean survival time (RMST) is the mean survival time in the study population
followed up to a specific time point, 
and is simply the area under the survival curve up to the specific time point.
The difference between two RMSTs quantifies the group difference in a time scale by measuring 
the integrated difference between survival curves in the two treatment groups. 
This paper develops a group sequential (GS) test of comparing
two RMSTs up to a restriction time point under a stratified proportional hazards model
with stratum representing treatment status. 
This covariate-adjusted GS test does not require the assumption of a constant hazard ratio
over time between the two treatment groups and is valid 
whether or not the proportional hazards assumption holds for the treatment effect. 
We establish the large-sample properties of the covariate-adjusted GS test 
and show that the proposed test statistics sequentially computed at
different interim analysis times possess an independent increments covariance structure.
Using currently available methodology, this joint independent increments structure 
allows us to calculate sequential stopping boundaries for 
preserving the desired type I error probability and maintaining 
the required statistical power. 
We evaluate the small-sample performance of the proposed GS
test via a simulation study and illustrate 
its real-world application through analysis of a clinical trial dataset from the BMT CTN 1101 study.
\end{abstract}


\section{Introduction}

\label{sec1}
The survival probability at a specific event time point $\tau$ describes the chance that 
an individual survives longer than $\tau$ units of time. 
This summary measure for time-to-event outcomes, however, does
not contain information regarding the temporal profile
of the survival time up to $\tau$. 
An alternative yet useful summary measure is the
restricted mean survival time (RMST) at time $\tau$,
also called $\tau$-year mean survival time,
which is the mean survival time of all subjects
in the study population followed up to the
specific time point $\tau$. Selection of the appropriate clinically relevant time horizon can be challenging and should be done in consultation with clinicians or investigators.  
The restricted time is often considered, 
as information on patient outcomes may not be available
beyond a certain time point due to limited study follow-up. 
The RMST summarizes the temporal event time distribution
during the time interval [0, $\tau$] and corresponds
to the area under the survival curve up to
the survival time $\tau$. 
The RMST is an alternative summary measure to the survival function
and provides information beyond the
survival probability only at a single time point. 

Covariate-adjusted analyses are often used in randomized clinical trials;
the use of adjustment for important prognostic factors may improve
estimation precision and increase power to detect a between-group treatment effect.
The adjusted analysis is also helpful in accounting for observed chance imbalances
in distribution of some baseline covariates, 
even though randomization ensures comparability among treatment arms on average. 

When comparing survival curves between two treatment groups, 
the hazard ratio (HR) under the Cox proportional hazards model 
is a frequently used effect measure in survival analysis to evaluate 
the treatment effect. 
The clinical and statistical interpretation of survival benefit using HRs
is appropriate when the proportional hazards (PH) assumption
holds. 
However, in the case of non-proportional hazards (NPH), 
when the PH assumption is violated,
or when it is difficult to ascertain whether the PH assumption holds,
it may be misleading to evaluate the treatment effect
using the HR in that reporting only the HR may lead to invalid conclusions. 
Survival curves having NPH can occur when a therapy has a diminishing effect
or a delayed treatment effect. 
As pointed out by \cite{Chen13}, immunotherapies in oncology often have a 
delayed treatment effect when compared to chemotherapy or radiation therapy.
An alternative measure that can be employed to quantify
the difference of two treatment groups is the difference between
their RMSTs up to a pre-specified time point. 
The RMST is informative and clinically interpretable for assessing
the effect of a treatment between the treatment arms regardless of whether 
the PH assumption is valid. 
The use of RMST has been discussed recently in clinical 
literature as an alternative summary measure to the HR; see, for example \cite{RoystParm2013}, \cite{EatoTher20}, and \cite{Hase20}.
The RMST difference has two appealing advantages:
(i) it is easily interpretable as the mean difference in survival times
by the end of follow-up and (ii) it does not require
the proportional hazards assumption. 

Group sequential (GS) designs are commonly employed in clinical trials
with censored survival data and staggered entry,
in which data are examined periodically and 
treatment comparisons are performed repeatedly at successive analyses over time. 
These multiple interim evaluations of accumulating data
allow researchers to potentially declare early success or futility in clinical trials, and offer a potential reduction in the resources
and monetary costs of the trial. 
According to \cite{Todd2007}, the benefits of early termination of
a clinical trial are most substantial in the late stage of the trial,
such as phase III trials, where the sample sizes of trials at this stage are typically large.
Moreover, GS studies have an ethical advantage in that 
periodic data monitoring helps ensure that individuals are not 
exposed to unsafe, inferior, or ineffective treatments unnecessarily
when there is strong evidence showing that one treatment is superior
to the other at the time of an interim analysis.

Using the sequentially computed Kaplan-Meier survival estimates studied by \cite{JennTurn99},
\cite{MurrTsia99} studied nonparametric GS tests for comparing RMSTs,
where the RMST estimate for each arm is obtained by integrating its Kaplan-Meier curve.
But, these GS tests do not adjust for covariates that may be predictive 
of survival outcomes. Our work in this paper is motivated by the Blood
and Marrow Transplant Clinical Trials Network (BMT CTN) trial 1101 
described by \cite{Fuchs2020}, 
which compared post-transplant survival outcomes between two treatment arms, 
double umbilical cord blood and HLA-haploidentical bone marrow transplantation, in the presence of nine baseline covariates. 
The design of this randomized multicenter phase 3 trial took into 
consideration concerns about the possibility of NPH.
Since the assumption of a constant hazard ratio over time between the
two groups may not hold in practice, we consider in this paper
covariate-adjusted GS comparison of RMSTs that avoids the PH assumption.
This GS procedure employs covariate-adjusted treatment specific 
survival function estimates under a treatment-stratified Cox PH 
regression model discussed by \cite{Zuck98, Zhan13} and uses their difference
as the basis of comparison between the treatment arms. 
Two attractive features of the proposed GS study of treatment comparisons
using covariate-adjusted RMSTs are (i) it does not require the 
proportionality of the hazards at each interim analysis when sequentially fitting the stratified PH model and 
(ii) it has a straighforward interpretation as the mean difference 
in survival times by the end of followup. 
If we use covariate-adjusted survival function
estimates under the Cox PH model, the resulting GS analysis of treatment
comparisons via covariate-adjusted RMSTs is only valid when the HR
between the two treatment groups is constant over time. 
As a useful alternative to the logrank and hazard ratio approaches for
time-to-event outcomes in a GS setting, 
the proposed GS test for comparing RMSTs between treatment arms, 
which appropriately adjusts for covariate imbalances and accounts for NPH, will lead to more efficient clinical trials. 

When repeatedly looking at the trial data and sequentially performing multiple hypothesis tests during the course of a GS study,
care must be taken to ensure the overall type I error probability is not inflated. \cite{SchaTsia97} demonstrated that an analysis method which is semiparametric efficient for the target parameter must yield sequential test statistics that possess independent increments asymptotically. The stratified Cox model is known to be semiparametric efficient for estimating its log hazard ratio parameters \citep{ZengLin06}. However, because the parameter of interest in the method of \cite{Zhan13} is the difference in adjusted RMSTs, each of which is a function of both the log hazard parameters and the infinite dimensional, unspecified baseline hazard function of the Cox model, these findings do not imply that the test statistics from adjusted RMST differences will have independent increments in a GS test. We instead directly derive the large sample distribution of these test statistics using martingale theory, showing that they have independent increments asymptotically. This permits readily available methods to be employed for determining critical values that satisfy type I error rate and power specifications. \cite{VanBetz25} proposed an iterative orthogonalization method to update an initial sequence of estimators to a modified sequence of estimators that possesses asymptotically the independent increments structure and has an asymptotic variance no larger than that of the initial sequence of estimators. If the initial sequence of estimators already possesses the independent increments structure, the initial sequence of estimators and the modified sequence of estimators are asymptotically identical and have equal asymptotic variance matrices, so the orthogonalization approach adds unneeded complexity and superfluous computations. If the initial sequence of estimators truly follows the independent increments structure, this fact still needs to be proved directly as it is not a consequence of Theorem 1 of \cite{VanBetz25}. Since we have derived that the test statistics from the proposed GS test of the RMST difference has the independent increments structure, the orthogonalization approach is not necessary and using the proposed GS test would lead to more direct and straightfoward calculations for researchers and clinicians. \cite{Diaz19} proposed a covariate-adjusted estimator of the RMST in randomized trials with discrete survival outcomes without the PH assumption. However, their estimator utilizes the nonparametric boostrap to obtain an estimator for its asymptotic variance. In this paper, we have derived a closed form expression for a consistent variance estimator of the proposed covariate-adjusted RMST difference estimator, which is computationally more efficient than a nonparametric bootstrap. Note also that the pseudo-value approach, discussed in \cite{Ande03}, \cite{Ande04}, and \cite{Loga11}, has been used for covariate adjusted comparisons of RMSTs, but it has not been studied in the group sequential setting and we do not consider it further here.  

In order to determine the sequential boundaries 
for preserving the overall type I error probability at
the prespecified level and maintain statistical power at the required level, 
we derive the asymptotic joint distribution of our covariate-adjusted test statistics sequentially computed at different analysis times. 
This joint distribution is shown to possess an independent increments structure, which allows us to
perform the proposed GS tests using standard stopping methods for GS clinical trials, including Pocock and O
Brien-Fleming designs \citep{Poco77, OBriFlem79} and error spending functions \citep{LanDeMe83}.
Since the asymptotic distribution approximates the joint distribution of repeatedly computed GS test statistics and is guaranteed to hold only in large samples, we will perform simulation studies of the proposed GS tests to gain an understanding of their operating characteristics for small studies. 
Finally, we will apply the proposed covariate-adjusted RMST GS tests to a real clinical trial dataset from the aforementioned BMT CTN 1101 study. 

This paper is organized as follows. Section 2 proposes a GS test for treatment comparisons via covariate-adjusted RMSTs and
studies the asymptotic behaviors of repeatedly computed test statistics over time.
Section 3 presents a simulation study that investigates 
the finite-sample performance of the proposed GS test in terms of meeting type I error rate
and power specifications for a randomized group sequential trial. 
Section 4 applies the covariate-adjusted test to a real clinical trial dataset from the BMT CTN 1101 study.
A discussion of these findings is included in Section 5. Theoretical derivations supporting this methodology are presented in the Appendix.

\section{Methods}
\subsection {Data Notation}

Consider a typical randomized clinical trial with a survival endpoint and staggered enrollment
that compares a treatment group with a control group.
Under a GS design for time-to-event data, since individuals enter the trial at staggered intervals,
there are two time scales to be considered: the survival time denoted by $t$ and the calendar time denoted by $u$. Let the treatment and control groups be indexed by $i = 1$ and $i = 0$, respectively, and let $n_i$ be the number of patients in group $i$.
For the $j$th individual in group $i$,
let $E_{ij}$ be the calendar time of enrollment, $T_{ij}$ the time from entry to event,
$C_{ij}$ the time from entry to right censoring,
$\bfZ_{ij} = (Z_{ij1}, \ldots, Z_{ijp})^T$ a $p$-dimensional column vector of baseline covariates,
and $n=n_0+n_1$ be the total sample size. Assume an upper bound $u^*$ exists on the calendar age of the study.

We assume throughout this paper that
the quadruples $\{(E_{ij}, T_{ij}, C_{ij}, \bfZ_{ij}),~ j=1, \ldots, n_i\}$
are independent and identically distributed for $i=0,1$
and that the random vectors \\ $\{(E_{0j}, T_{0j}, C_{0j}, \bfZ_{0j}),~ j=1, \ldots, n_0\}$ and
$\{(E_{1j}, T_{1j}, C_{1j}, \bfZ_{1j}),~ j=1, \ldots, n_1\}$ are independent. Conditionally independent censoring is assumed such that $T_{ij} \perp C_{ij} | \mathbf{Z}_{ij}, I(\mbox{treatment = } i)$ for all $i, j$. 
The observed data in a group sequential study with right-censored failure time data and staggered enrollment
are composed of
independent observations $\{X_{ij}(u), \delta_{ij}(u), \bfZ_{ij}, i = 0,1, j = 1, \ldots, n_i\}$,
where $X_{ij}(u) = \min\{T_{ij}, C_{ij}, (u-E_{ij})^+\}$ 
is the observed time-on-study at analysis time $u$,
$\delta_{ij} = I(T_{ij} \leq \min \{C_{ij}, (u-E_{ij})^+\})$
is the indicator of the event that failure has occurred by calendar time $u$,
and $a^+= \max\{0, a\}$ for a real number $a$.

\subsection {Covariate-adjusted RMSTs}

We aim to develop a two-sample GS covariate adjusted comparison of RMSTs between the treatment and control arms by extending the Cox model-adjusted RMSTs of \cite{Zuck98} and \cite{Zhan13}
that were previously studied in the fixed sample design setting.
To adjust this treatment comparison for the effects of baseline covariates,
we adopt the following treatment-stratified Cox proportional hazards regression model \citep{Kalb11}
in which the conditional hazard function of $T_{ij}$ given $\bfZ$ is modeled as
\be
\label{eq:stratcox2}
\lambda_i(t|\bfZ) = \lambda_{0i}(t) \exp({\boldsymbol\beta}^T\bfZ) ~~
\hbox{for treatment arm}~ i=0,1 ~~\hbox{and}~~ 0 \leq t \leq \tau,
\ee
where $\lambda_{0i}(t)$ is an unknown treatment-specific baseline hazard function of $T_{ij}$
for treatment arm $i$ and $\bfbeta = (\beta_1, \ldots, \beta_p)^T$ 
is a $p \times 1$ vector of parameters to be estimated.
Under model (\ref{eq:stratcox2}), 
the difference in RMSTs between the two groups measures the treatment effect with covariate adjustment
and is an alternative to the HR to quantify the effect of an intervention
without relying on the proportionality of the hazards between treatment groups.
For $i=0,1$, the treatment-specific baseline cumulative hazard and survival functions 
are, respectively, equal to $\Lambda_{0i}(t) = \int_0^t \lambda_{0i}(s)ds$ and $S_{0i}(t) = \exp\{-\Lambda_{0i}(t)\}$,
where $\lambda_{0i}(t)$ is assumed to be continuous.

Let $\tau > 0$ be a specific time point of interest.
The RMST up to time $\tau$ for treatment arm $i$
is defined as the expected value of the minimum of $T_{ij}$ and $\tau$, i.e.
\be
\mu_i(\tau) = E\{\min(T_{ij}, \tau)\} = \int_0^{\tau} S_i(t)dt, \qquad i=0,1,
\ee
where $S_i(t)=P(T_{ij} \geq t)$ is the survival function of $T_{ij}$
and $\min(T_{ij}, \tau)$ is the survival time  restricted to $[0,\tau]$.
The mean survival time $\mu_i(\tau)$ within the specific time window $[0, \tau]$
is often referred to as the $\tau$-year mean survival time
and corresponds visually to the area under the curve of the survival function $S_i(t)$ over $[0, \tau]$. As an alternative to the hazard ratio to quantify the effect of an intervention,
we can perform the treatment group comparison based on the RMSTs of the groups.
The difference between treatment arms in RMSTs is given by
$\Delta(\tau) = \mu_1(\tau)-\mu_0(\tau) 
= \int_0^{\tau} \{S_1(t)-S_0(t)\}dt,$
which can be regarded as a summary measure to quantify the survival benefit of the treatment over $[0,\tau]$.

Under model \ref{eq:stratcox2},
the conditional cumulative hazard function and conditional survival function of $T_{ij}$ given $\bfZ=\bfz$
are, respectively, calculated as
$\Lambda_i(t|\bfz) = \int_0^t \lambda_i(s|\bfz)ds = \exp(\bfbeta_0^T\bfz)\Lambda_{0i}(t)$
and $S_i(t|\bfZ) = P(T_{ij} \geq t|\bfZ=\bfz) = \exp\{-\Lambda_i(t|\bfz)\} = \{S_{0i}(t)\}^{\exp(\bfbeta_0^T\bfz)}$,
where $\bfbeta_0$ is the true value of $\bfbeta$.
\cite{Zuck98} defined the following average survival function for group $i$,
\Eq{align}
{
\label{eq:Sitilde}
\tilde S_i(t) = {1 \over n} \sum_{g=0}^1\sum_{j=1}^{n_g} S_i(t|\bfZ_{gj})
= {1 \over n} \sum_{g=0}^1\sum_{j=1}^{n_g} \exp\{-\exp(\bfbeta_0^T\bfZ_{gj}) \Lambda_{0i}(t)\},
}
where the averaging is taken over all the observed covariate vectors $\{\bfZ_{ij}\}$ from the two groups.
Since $E\{\tilde S_i(t)\} = E\{S_i(t|\bfZ)\}=S_i(t)$ for $i=0,1$,
the survival function $S_i(t)$ can be regarded as an average survival curve for group $i$.

Similarly, the average restricted mean survival time is defined as\Eq{align}
{
\label{eq:muitilde}
\tilde\mu_i(\tau) &= \int_0^{\tau} \tilde S_i(t)dt
= \int_0^{\tau} {1 \over n} \sum_{g=0}^1\sum_{j=1}^{n_g} S_i(t|\bfZ_{gj}) dt
= {1 \over n} \sum_{g=0}^1\sum_{j=1}^{n_g} \mu_i(\tau|\bfZ_{gj}) \cr
&= {1 \over n} \sum_{g=0}^1\sum_{j=1}^{n_g} \int_0^{\tau}
\exp\{-\exp(\bfbeta_0^T\bfZ_{gj}) \Lambda_{0i}(t)\} dt, \qquad i=0,1,
}
where $\mu_i(\tau|\bfZ) = \int_0^{\tau} S_i(t|\bfZ)dt$ is the conditional treatment-specific
restricted mean survival time.
The average restricted mean survival times $\tilde\mu_0(\tau)$ and $\tilde\mu_1(\tau)$ give rise to
the difference between treatment arms in average restricted mean survival times,
$\tilde \Delta(\tau) = \tilde\mu_1(\tau)-\tilde\mu_0(\tau)
= \int_0^{\tau} \{\tilde S_1(t)-\tilde S_0(t)\}dt,$
which can be alternatively used as a summary measure to quantify survival benefit of the treatment as proposed in \cite{Zuck98}.
Since $E\{\tilde S_i(t)\} = E\{S_i(t|\bfZ)\}=S_i(t)$, it is seen from \ref{eq:muitilde} that
\bes
\ba
& E\{\tilde\mu_i(\tau)\} 
= {1 \over n} \sum_{g=0}^1\sum_{j=1}^{n_g} \int_0^{\tau} E\{S_i(t|\bfZ_{gj})\} dt
= {1 \over n} \sum_{g=0}^1\sum_{j=1}^{n_g} \int_0^{\tau} S_i(t) dt
= \mu_i(\tau) \cr
\ea
\ees
and $E\{\tilde\Delta(\tau)\} = E\{\tilde\mu_1(\tau)\}-E\{\tilde\mu_0(\tau)\}=\mu_1(\tau)-\mu_0(\tau) = \Delta(\tau)$.
In the next section,
we study group sequential testing of the RMST difference $\Delta(\tau)$, the parameter quantifying the treatment effect.

\subsection {Estimation of Covariate-adjusted RMSTs under Staggered Entry}

Using the counting process representation for the partial score function
of the Cox proportional hazards regression model under staggered entry,
\cite{Bila97} defined a two-parameter partial score process
indexed by both the calendar time and the survival time.
Similarly, we can consider a two-parameter partial score process
under the treatment-stratified Cox proportional hazards regression model \ref{eq:stratcox2}.
At respective calendar and survival times $u$ and $t$,
the two-parameter partial score process based on data available up to
a specified survival time $t \in [0,\tau]$ is defined as
\Eq{align*}
{
\mathbf{U}(\bfbeta,u,t)
= \sum_{i=0}^1 \sum_{j=1}^{n_i} \int_0^t \bigl\{\bfZ_{ij} - \mathbf{E}_i(\bfbeta,u,s)\bigr\} N_{ij}(u,ds).
}
and the partial observed information matrix is
\Eq{align*}
{
\mathbf{I}(\bfbeta,u,t) 
= \int_0^t \sum_{i=0}^1 \sum_{j=1}^{n_i} \mathbf{V}_i(\bfbeta,u,s) N_{ij}(u,ds),
}
where $N_{ij} (u,t) = I\bigl(X_{ij}(u) \leq t, \delta_{ij}(u) = 1\bigr)$
is the counting process of observed failures in $(0, t]$ 
for the $j$th individual in group $i$ at analysis time $u$,
$Y_{ij} (u,t) = I\bigl(X_{ij}(u) \geq t\bigr)$ is the corresponding at-risk process, and 
\Eq{align*}
{
& S_{0i}(\bfbeta,u,t) = {1 \over n_i} \sum_{j=1}^{n_i} Y_{ij}(u,t) \exp(\bfbeta^T \bfZ_{ij}), \
\mathbf{S}_{1i}(\bfbeta,u,t) = {1 \over n_i} \sum_{j=1}^{n_i} Y_{ij}(u,t) \exp(\bfbeta^T \bfZ_{ij})\bfZ_{ij}, \cr
& \mathbf{S}_{2i}(\bfbeta,u,t) = {1 \over n_i} \sum_{j=1}^{n_i} Y_{ij}(u,t) \exp(\bfbeta^T \bfZ_{ij})\bfZ_{ij}\bfZ_{ij}^T, \cr
& \mathbf{E}_i(\bfbeta,u,t) = {\mathbf{S}_{1i}(\bfbeta,u,t) \over S_{0i}(\bfbeta,u,t)}, \qquad
\mathbf{V}_i(\bfbeta,u,t) = {\mathbf{S}_{2i}(\bfbeta,u,t) \over S_{0i}(\bfbeta,u,t)} - \mathbf{E}_i(\bfbeta,u,t) \mathbf{E}_i^T(\bfbeta,u,t).
}
The expected partial information matrix is found to be
$\cI(\bfbeta,u,t) = E\{\mathbf{I}(\bfbeta,u,t)\} = \Var\{\mathbf{U}(\bfbeta,u,t)\}.$

Define $\mathcal{D} = \{(u,t): 0 \le t \le u \le u^*, 0 \le t \le \tau\}$. For calendar and survival times $(u,t) \in \mathcal{D}$,
let $\hat\bfbeta(u,t)$ denote the maximum partial likelihood estimator (MPLE) of $\bfbeta$ under model \ref{eq:stratcox2}.
If the true value $\bfbeta_0$ of $\bfbeta$ were known,
we can employ the standard \cite{Bres74} procedure to estimate
the treatment-specific baseline cumulative hazard function $\Lambda_{0i}(t)$ at calendar time $u$ by
$\hat\Lambda_{0i}(\bfbeta_0,u,t)$, where
\Eq{align}
{
\label{eq:cumhaz}
\hat\Lambda_{0i}(\bfbeta,u,t)
&= \int_0^t {\sum_{j=1}^{n_i} N_{ij}(u,ds) \over \sum_{j=1}^{n_i} Y_{ij}(u,s) \exp(\bfbeta^T\bfZ_{ij})}
= {1 \over n_i} \sum_{j=1}^{n_i} \int_0^t {N_{ij}(u,ds) \over S_{0i}(\bfbeta,u,s)}.
}
When $\bfbeta_0$ is unknown,
it is natural to estimate $\Lambda_{0i}(t)$ at calendar time $u$ by the Breslow estimator
$\hat\Lambda_{0i}\bigl(\hat\bfbeta(u,t),u,t\bigr)$.
Note that
$\mathbf{U}(\bfbeta,u,t)$, $\hat\bfbeta(u,t)$, and $\hat\Lambda_{0i}\bigl(\hat\bfbeta(u,t),u,t\bigr)$
are two-parameter stochastic processes indexed by $(u,t)$.

To compare the two treatment groups in terms of the difference of their RMSTs,
we extend covariate-adjusted restricted mean survival times \citep{Zuck98, Zhan13} 
to the group sequential setting under model \ref{eq:stratcox2} for testing the null hypothesis $H_0: \mu_0(\tau)=\mu_1(\tau)$
versus the two-sided alternative hypothesis $H_1: \mu_0(\tau) \not= \mu_1(\tau)$
or the one-sided alternative hypothesis $H_1: \mu_0(\tau) < \mu_1(\tau)$ or $H_1: \mu_0(\tau) > \mu_1(\tau)$.

In order to make use of all available data at calendar time $u$ to estimate $\boldsymbol{\beta}$ from model \ref{eq:stratcox2}, the estimator $\hat{\boldsymbol{\beta}}(u,\min\{u,\tau\})$ should be used. For the remainder of this section, we assume this is done for all $u$ according to $\hat{\boldsymbol{\beta}}(u) = \hat{\boldsymbol{\beta}}(u,\min\{u,\tau\})$, suppressing the second argument.  
For $(u,t) \in \mathcal{D}$,
since $E\{\tilde S_i(t)\}=S_i(t)$, $E\{\tilde\mu_i(\tau)\}=\mu_i(\tau)$, and $E\{\tilde\Delta(\tau)\} = \Delta(\tau)$,
it is natural to estimate $S_i(t)$, $\mu_i(\tau)$, and $\Delta(\tau)$ at calendar time $u$ by
\Eq{align*}
{
& \hat S_i\bigl(\hat\bfbeta(u);u,t\bigr)
= {1 \over n} \sum_{g=0}^1\sum_{j=1}^{n_g}
\exp[-\exp\{\hat\bfbeta^T(u)\bfZ_{gj}\} \hat\Lambda_{0i}(\hat\bfbeta(u);u,t)], \cr
& \hat\mu_i(u,\tau) = \int_0^{\tau} \hat S_i\bigl(\hat\bfbeta(u);u,t\bigr)dt
= {1 \over n} \sum_{g=0}^1\sum_{j=1}^{n_g}
\int_0^{\tau} \exp[-\exp\{\hat\bfbeta^T(u)\bfZ_{gj}\} \hat\Lambda_{0i}(\hat\bfbeta(u);u,t)]dt, \cr
& \hat\Delta(u,\tau) = \hat\mu_1(u,\tau)-\hat\mu_0(u,\tau)
= \int_0^{\tau} \bigl\{\hat S_1\bigl(\hat\bfbeta(u);u,t\bigr)
- \hat S_0\bigl(\hat\bfbeta(u);u,t\bigr)\bigr\}dt,
}
where $\hat\Lambda_{0i}\bigl(\hat\bfbeta(u);u,t)$ is the Breslow estimator of $\Lambda_{0i}(t)$ 
at calendar time $u$ with
$\hat\Lambda_{0i}(\bfbeta,u,t)$ defined in \ref{eq:cumhaz}.
Thus, under model \ref{eq:stratcox2},
the RMSTs $\mu_0(\tau)$ and $\mu_1(\tau)$ are estimated for each treatment arm at calendar time $u$
by finding the area up to time $\tau$
under the adjusted survival curve $\hat S_i\bigl(\hat\bfbeta(u);u,t\bigr)$.
The difference between the treatment-specific RMST estimators $\hat\mu_i(u,\tau)$
leads to an estimator of $\Delta(\tau)$ and can be employed as the basis of a test statistic
of $H_0: \mu_0(\tau)=\mu_1(\tau)$.

For a calendar time $u$, let $t_{i(1)} < \cdots < t_{i(r_i)} \leq u$, $i = 0, 1$, denote the ordered failure times in a sample of size $n_i$ from treatment group $i$ with $t_{i(0)}=0$ and $t_{i,(r_i+1)}=\tau$,
where $r_i$ is the number of distinct failures observed in group $i$ over $[0, \tau]$.
Suppose that $d_{ik}$ failures occur at $t_{i(k)}$ and that $y_{ik}$ study subjects are
at risk just prior to $t_{i(k)}$ for $i=0,1$ and $k = 1, \ldots, r_i$.
Then the treatment-specific covariate-adjusted RMST estimator $\hat\mu_i(u,\tau)$ can be expressed as
\Eq{align*}
{
& \hat\mu_i(u,\tau) 
= {1 \over n} \sum_{g=0}^1\sum_{j=1}^{n_g} \sum_{k=1}^{r_i}
\exp[-\exp\{\hat\bfbeta^T(u)\bfZ_{gj}\} \hat\Lambda_{0i}(\hat\bfbeta(u);u,t_{i(k)})](t_{i(k+1)}-t_{i(k)}),
}
from which we can calculate the covariate-adjusted estimator $\hat\Delta(u,\tau)$ of the RMST difference as
\Eq{align*}
{
& \hat\Delta(u,\tau) = \hat\mu_1(u,\tau)-\hat\mu_0(u,\tau) \cr
&~~= {1 \over n} \sum_{g=0}^1\sum_{j=1}^{n_g} \biggl(\sum_{k=1}^{r_1}
\exp[-\exp\{\hat\bfbeta^T(u)\bfZ_{gj}\} \hat\Lambda_{01}(\hat\bfbeta(u);u,t_{1(k)})](t_{1(k+1)}-t_{1(k)}) \cr
&\qquad - \sum_{k=1}^{r_0}
\exp[-\exp\{\hat\bfbeta^T(u)\bfZ_{gj}\} \hat\Lambda_{00}(\hat\bfbeta(u);u,t_{0(k)})](t_{0(k+1)}-t_{0(k)})\biggr).
}

\subsection {Large-sample Properties of $\hat\Delta(u,\tau)$}
Let $\mathcal{D}_* = \{(u,t): u_* \le t \le u \le u^*, t \le \tau\}$, where $u_*$ is a calendar time such that $\boldsymbol{\Sigma}(\boldsymbol{\beta_0},u)$ is positive definite for $u_* \le u \le u^*$. In this section, we study the asymptotic behavior of the stochastic process
$\bigl\{\sqrt{n}\hat\Delta(u,\tau),~ u_* \leq u \leq u^* \bigr\}$
under the null hypothesis $H_0: \mu_0(\tau)=\mu_1(\tau)$.
Theorems 1 and 2 establish the asymptotic distributions of the stochastic processes
$\bigl\{\sqrt{n} \{\hat\Delta(u,\tau)-\tilde\Delta(u,\tau)\},~ u_* \leq u \leq u^* \bigr\}$ and
$\bigl\{\sqrt{n} \{\hat\Delta(u,\tau)-\Delta(u,\tau)\},~ u_* \leq u \leq u^* \bigr\}$, respectively.

\setcounter{theorem}{0}
\begin{theorem}
Suppose that the regularity conditions (A)-(H) in the Appendix are satisfied for $u \in [u_*, u^*]$.
Then under the stratified proportional hazards model \ref{eq:stratcox2},
the stochastic process
$\bigl\{\sqrt{n} \{\hat\Delta(u,\tau)-\tilde\Delta(u,\tau)\},~ u_* \leq u \leq u^* \bigr\}$
converges in distribution to a Gaussian random process $\xi$
with mean 0 and covariance function specified by
\Eq{align*}
{
& V_{\xi}(u_1,u_2,\tau) = E\{\xi(u_1,\tau)\xi(u_2,\tau)\} \cr
&\quad= {1 \over \rho_0} \int_0^{\tau}\int_0^{\tau}
c_{01}(s)c_{01}(t) \gamma_0(\bfbeta_0,u_1 \vee u_2, s \wedge t\bigr)dsdt \cr
&\qquad + {1 \over \rho_1} \int_0^{\tau}\int_0^{\tau} c_{11}(s)c_{11}(t) 
\gamma_1(\bfbeta_0,u_1 \vee u_2, s \wedge t\bigr)dsdt \cr
&\qquad + \biggl\{\int_0^{\tau} \mathbf{D}(\bfbeta_0,s)ds\biggr\}^T \boldsymbol{\Sigma}^{-1}(\bfbeta_0,u_1 \vee u_2)
\biggl\{\int_0^{\tau} \mathbf{D}(\bfbeta_0,t) dt\biggr\} \cr
&\quad= K_{\xi}(u_1 \vee u_2, u_1 \vee u_2,\tau) 
= \Var\{\xi(u_1 \vee u_2,\tau)\}, \qquad u_* \leq u_1, u_2 \leq u^*. \hskip0.4in
}
where $\rho_i$ and $\boldsymbol{\Sigma}(\bfbeta_0,u)$ are defined 
in Conditions (E) and (F)
and $c_{i1}(t)$, $\gamma_i(\bfbeta_0,u,t)$, $\mathbf{D}(\bfbeta_0,t)$
are defined in \ref{eq:expansion} from the Appendix, respectively.
The asymptotic variance of $\sqrt{n} \{\hat\Delta(u,\tau)-\tilde\Delta(u,\tau)\}$
is given by $V_{\xi}^2(u,\tau)=V_{\xi}(u,u,\tau)$ and is consistently estimated by
$$\hat V_{\xi}^2(u,\tau) 
= \hat B_{10}\bigl(u,\tau\bigr) + \hat B_{11}\bigl(u,\tau\bigr) + \hat B_3\bigl(u,\tau\bigr),$$
where, 
\Eq{align}
{
& \hat B_{1i}(u,\tau)
= {n \over n_i} \sum_{j=1}^{r_i} \sum_{k=1}^{r_i}
\hat c_{i1}(u,t_{i(j)})\hat c_{i1}(u,t_{i(k)}) \hat\gamma_i(u, t_{i(j)} \wedge t_{i(k)}\bigr)
(t_{i(j+1)}-t_{i(j)})(t_{i(k+1)}-t_{i(k)}), \cr
& \hat B_3(u,\tau)
= n\{\hat{\boldsymbol{\Psi}}_1(u,\tau)-\hat{\boldsymbol{\Psi}}_0(u,\tau)\}^T I^{-1}\bigl(\hat\bfbeta(u),u\bigr)
\{\hat{\boldsymbol{\Psi}}_1(u,\tau)-\hat{\boldsymbol{\Psi}}_0(u,\tau)\}, \cr
& \hat{\boldsymbol{\Psi}}_i(u,\tau)
= \sum_{j=1}^{r_i} \{\hat c_{i1}(u,t_{i(j)})\hat{\mathbf{Q}}_i(u,t_{i(j)})
-\hat\Lambda_{0i}\bigl(\hat\bfbeta(u);u,t_{i(j)}\bigr) \hat{\mathbf{c}}_{i2}(u,t_{i(j)})\} (t_{i(j+1)}-t_{i(j)}), \cr
& \hat c_{i1}(u,t) = {1 \over n} \sum_{g=0}^1\sum_{j=1}^{n_g}
\hat S_i\bigl(\hat\bfbeta(u);u,t|\bfZ_{gj}\bigr)\exp\{\hat\bfbeta(u)^T\bfZ_{gj}\}, \cr
& \hat{\mathbf{c}}_{i2}(u,t) = {1 \over n} \sum_{g=0}^1\sum_{j=1}^{n_g}
\hat S_i\bigl(\hat\bfbeta(u);u,t|\bfZ_{gj}\bigr)\exp\{\hat\bfbeta(u)^T\bfZ_{gj}\}\bfZ_{gj}, \cr
& \hat\gamma_i(u,t) = {1 \over n_i} \int_0^t {dN_i(u,s) \over S_{0i}^2\bigl(\hat\bfbeta(u);u,s\bigr)}, \qquad
\hat{\mathbf{Q}}_i(u,t)
={1 \over n_i} \int_0^t {\mathbf{S}_{1i}\bigl(\hat\bfbeta(u);u,s\bigr) \over 
S_{0i}^2\bigl(\hat\bfbeta(u);u,s\bigr)}dN_i(u,s).
}
\end{theorem}

\begin{theorem}
Suppose that the regularity conditions (A)-(H) in the Appendix are satisfied for $u \in [u_*, u^*]$.
Then under the stratified proportional hazards model \ref{eq:stratcox2},
the stochastic process
$\bigl\{\sqrt{n} \{\hat\Delta(u,\tau)-\Delta(u,\tau)\},~ u_* \leq u \leq u^* \bigr\}$
converges in distribution to a Gaussian random process $\eta$
with mean 0 and covariance function specified by
\Eq{align*}
{
& V_{\eta}(u_1,u_2,\tau) = E\{\eta(u_1,\tau)\eta(u_2,\tau)\} \cr
&\quad= V_{\xi}(u_1,u_2,\tau) + \Var\{\mu_1(\tau|\bfZ)-\mu_0(\tau|\bfZ)\} \cr
&\quad = V_{\xi}(u_1 \vee u_2, u_1 \vee u_2,\tau) + \Var\{\mu_1(\tau|\bfZ)-\mu_0(\tau|\bfZ)\},
\qquad u_* \leq u_1, u_2 \leq u^*. \hskip0.4in
}
For each $u \in [u_*,u^*]$,
the asymptotic variance of $\sqrt{n} \{\hat\Delta(u,\tau)-\Delta(u,\tau)\}$
is given by $V_{\eta}^2(u,\tau)=V_{\eta}(u,u,\tau)
= V_{\xi}(u,\tau) + \Var\{\mu_1(\tau|\bfZ)-\mu_0(\tau|\bfZ)\}$ and is consistently estimated by
\Eq{align}
{
\label{eq:thm2estimate}
& \hat V_{\eta}^2(u,\tau) 
= \hat V_{\xi}^2(u,\tau) + \widehat{\Var}\{\mu_1(\tau|\bfZ)-\mu_0(\tau|\bfZ)\},
}
where, for $i=0,1$,
\Eq{align*}
{
& \widehat{\Var}\{\mu_1(\tau|\bfZ)-\mu_0(\tau|\bfZ)\}  
= {1 \over n} \sum_{i=0}^1\sum_{j=1}^{n_i}
\{\hat\mu_1(\tau|\bfZ_{ij})-\hat\mu_0(\tau|\bfZ_{ij})-\hat\Delta(u,\tau)\}^2, \cr
& \hat\mu_i(\tau|\bfZ) = \int_0^{\tau} \hat S_i\bigl(\hat\bfbeta(u),u,t|\bfZ\bigr)dt
= \sum_{k=1}^{r_i}
\exp[-\exp\{\hat\bfbeta(u)^T\bfZ\} \hat\Lambda_{0i}(u,t_{i(k)})](t_{i(k+1)}-t_{i(k)}).
}
\end{theorem}
Proofs of Theorem 1 and Theorem 2 can be found in the Appendix. Theorems 1 and 2 imply that the asymptotic distribution of the RMST differences follow a reverse independent increments structure, which simplifies development of sequential tests of these differences by allowing the use of commonly used existing methods to calculate critical values.

\subsection{Group Sequential Testing of the Equality of RMSTs}
Let $u_1 < u_2 < \cdots < u_K$ denote the $K$ calendar times at which $K$ interim analyses are performed using the accumulating observed data corresponding to these calendar times.
The successive estimates $\hat\mu_i(u_1,\tau), \ldots, \hat\mu_i(u_K,\tau)$
of the treatment-specific RMST $\mu_i(\tau)$ with covariate adjustment under model \ref{eq:stratcox2}
allow for construction of group sequential tests of the equality of RMSTs.
Specifically,
a group sequential test of $H_0: \mu_0(\tau)=\mu_1(\tau)$ can be based on 
repeatedly computed Wald test statistics:
$$Z(u_k,\tau) = {\hat\Delta(u_k,\tau) \over \hat V_{\eta}(u_k,\tau)}
= {\hat\mu_1(u_k,\tau)-\hat\mu_0(u_k,\tau) \over \hat V_{\eta}(u_k,\tau)}, \qquad k=1, \ldots, K,$$
where $\hat V_{\eta}^2(u_k,\tau)$ from \ref{eq:thm2estimate} 
is a consistent estimate for the asymptotic variance $V_{\eta}^2(u_k,\tau)$ of
$\hat\Delta(u_k,\tau) = \hat\mu_1(u_k,\tau)-\hat\mu_0(u_k,\tau)$ at calendar time $u_k$.
The information level at analysis $k$ is given by $I(u_k,\tau) = \hat V_{\eta}^{-2}(u_k,\tau)$.
In group sequential monitoring of clinical trials,
a decision is made after each interim analysis to either stop the study due to early efficacy or to continue the trial on to the next analysis.
The reverse independent increments property derived in Theorem 2 enables the use of existing methods including Pocock and O’Brien-Fleming designs \citep{Poco77, OBriFlem79} and error spending functions \citep{LanDeMe83} to determine the rejection boundaries for sequential testing, 
preserving the overall type I error probability
and maintaining the required statistical power as closely as possible given a clinically meaningful RMST difference. For survival data,
error spending approaches are preferred due to the unpredictability of information increments over calendar times.

\section{Simulations}
Simulation studies were performed to: (1)  examine the proposed group sequential test's achieved type I error rate and power with reasonable sample sizes, and (2) contrast the achieved type I error rate and power of the proposed method to those of existing methods. 
We compared GS tests of the treatment effect based on the difference of the Cox model-adjusted RMSTs under the proposed method, the difference in nonparametric RMSTs computed using the area under the treatment groups' Kaplan-Meier curves \citep{MurrTsia99}, and the treatment effect's hazard ratio estimate in a
Cox PH model \citep{Bila97}. 
Randomized group sequential trials were simulated to study
the treatment effect between a treatment group and a control group by
testing the null hypothesis $H_0: \mu_0(\tau) = \mu_1(\tau)$ for both the proposed test and Kaplan-Meier based RMST test, and testing
$H_0^*: \beta_W = 0$ for the Cox model based test, where $\beta_W$ is the coefficient for
the treatment indicator $Z_W = I(\mbox{group = treatment})$.

Subjects were randomized between the treatment and control groups with a 1:1 allocation ratio.
They were uniformly enrolled during an accrual period [0, A];
let the length of the study be $L = \tau + A$.  
Group sequential tests were performed  
using the proposed test with three interim analyses,
and used the alpha spending function $0.05\min{\{1, IF^3}\}$ 
with a targeted significance level of 5\%, 
where $IF$ is the fraction of total information, $I_{Max}$, for the trial.
The expected information fractions at each analysis were set at $0.50, 0.75$, and $1$.
For each simulation scenario, the total information $I_{Max}$ and the calendar times for the three interim analyses that yielded the information fractions of $0.50, 0.75$, and $1$  were computed using Monte Carlo simulation.

Trial data were simulated using the following assumptions. 
Assume the event time for a subject follows a Weibull distribution with survival function
$S(t|Z_W, \mathbf{Z}) = \exp(-\gamma t^{\alpha})$, 
where the shape and rate parameters $\alpha$ and $\gamma$ depend on $Z_W$ and $\mathbf{Z}$:
$\alpha = \alpha_0 + \alpha_1Z_W$,
 $\gamma = \gamma_0 \exp(\beta_WZ_W + \bfbeta^T \bfZ)$. 
The hazard function for the Weibull distribution is
$\lambda(t | Z_W, \bfZ) =\alpha\gamma t^{\alpha - 1} = (\alpha_0 + \alpha_1Z_W) \gamma_0 \exp{(\beta_WZ_W + \bfbeta^T \bfZ)}t^{\alpha_0 + \alpha_1Z_W - 1}$.     
Then, the hazard ratio of the treatment groups equals $\lambda(t | Z_W = 1, \bfZ) /  \lambda(t | Z_W = 0, \bfZ) = (\alpha_0 + \alpha_1) / \alpha_0 \cdot \exp{(\beta_W)}t^{\alpha_1}$. 
Note that when $\alpha_1 =0$, the hazard ratio is $\exp{(\beta_W)}$, 
meaning that there are PH in this scenario;
when $\alpha_1 \not=0$, the PH assumption no longer holds.

Simulation parameters were chosen as follows:
 equal sample sizes per treatment group of $n_0 = n_1 = 200$ or 400;
 $\tau = 1$ or 1.5 years; 
independent censoring following an $\exp(-\log(0.95))$ distribution to yield 5\% censoring per year, or no censoring;
accrual period (0, A) where A = 2 or 4. 
Under $H_0$, $\beta_{W}$ was set to $\beta_{W0}$ by numerically solving for the value of $\beta_{W0}$ such that $\mu_1(\tau) - \mu_0(\tau) =0$. 
Under the alternative hypothesis $H_1: S_0(\tau) = S_1(\tau) - \delta$, $\beta_{W}$ was set to $\beta_{\delta}$,
the value of $\beta_{W}$ such that $\delta = \mu_1(\tau) - \mu_0(\tau)$, the effect size at which the proposed test is expected to achieve 80\% power.
For the influential covariates, $\mathbf{Z}$, we considered two possible distributions: a single covariate following the standard normal distribution,  and two covariates following standardized Bernoulli(0.3) and Bernoulli(0.5) distributions. 
For each covariate distribution, the covariate effects were specified as $\beta_j = \phi / \sqrt{p}$, where $p$ is the number of covariates and $\phi$ is a parameter that specifies the amount of covariate influence. Using this configuration, the variance of the linear predictor $\bfbeta^T \bfZ$ is equivalent under both covariate distributions for a given value of $\phi$. $\phi$ was chosen to be $0$, $\log(1.5)$, and $\log(2)$.
With these parameter settings, there are 96 possible parameter specifications.
We performed 2,000 simulated trials for each parameter specification and obtained Monte Carlo estimates to study the type I error rate and power of the proposed tests.

\begin{figure}[h]
 \caption{Cumulative Type I Error Rate and Power, NPH Setting. Dashed lines indicate nominal stagewise type I error rate and power values under asymptotic distribution.}
  \centering
  \includegraphics[width=8cm]{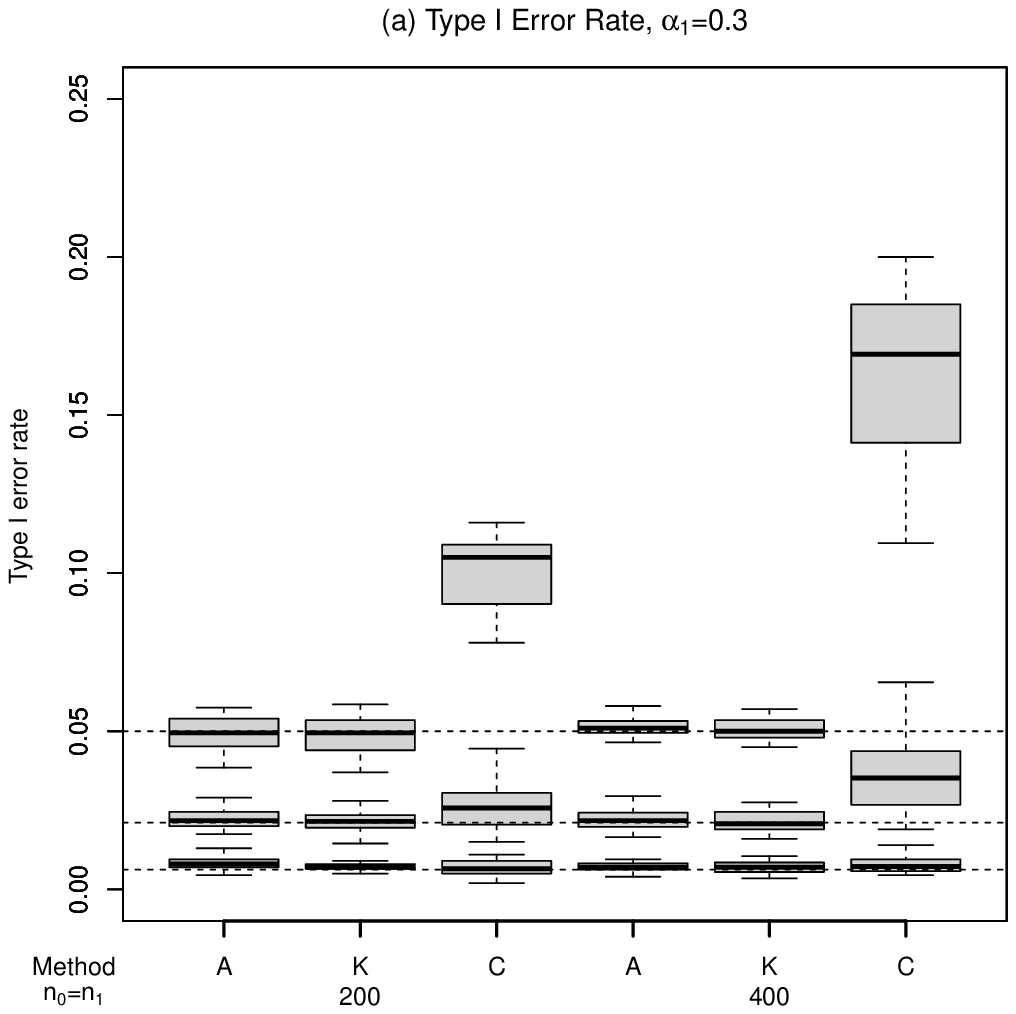}
  \includegraphics[width=8cm]{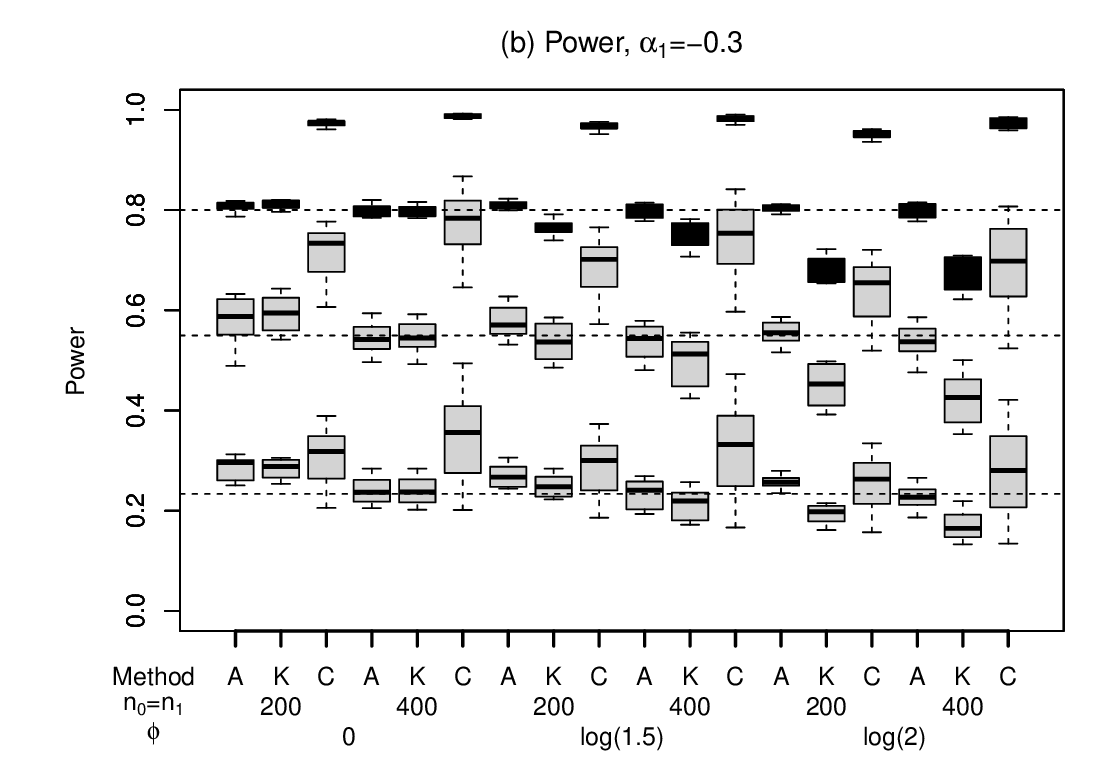}
  \vfill
   { \footnotesize A - RMST, 
  K - Kaplan-Meier based RMST, C - Cox Model}
\label{fig:3.1}
\end{figure}

\begin{figure}[h]
 \caption{Delayed Effect Curves}
  \centering
  \includegraphics[width=8cm]{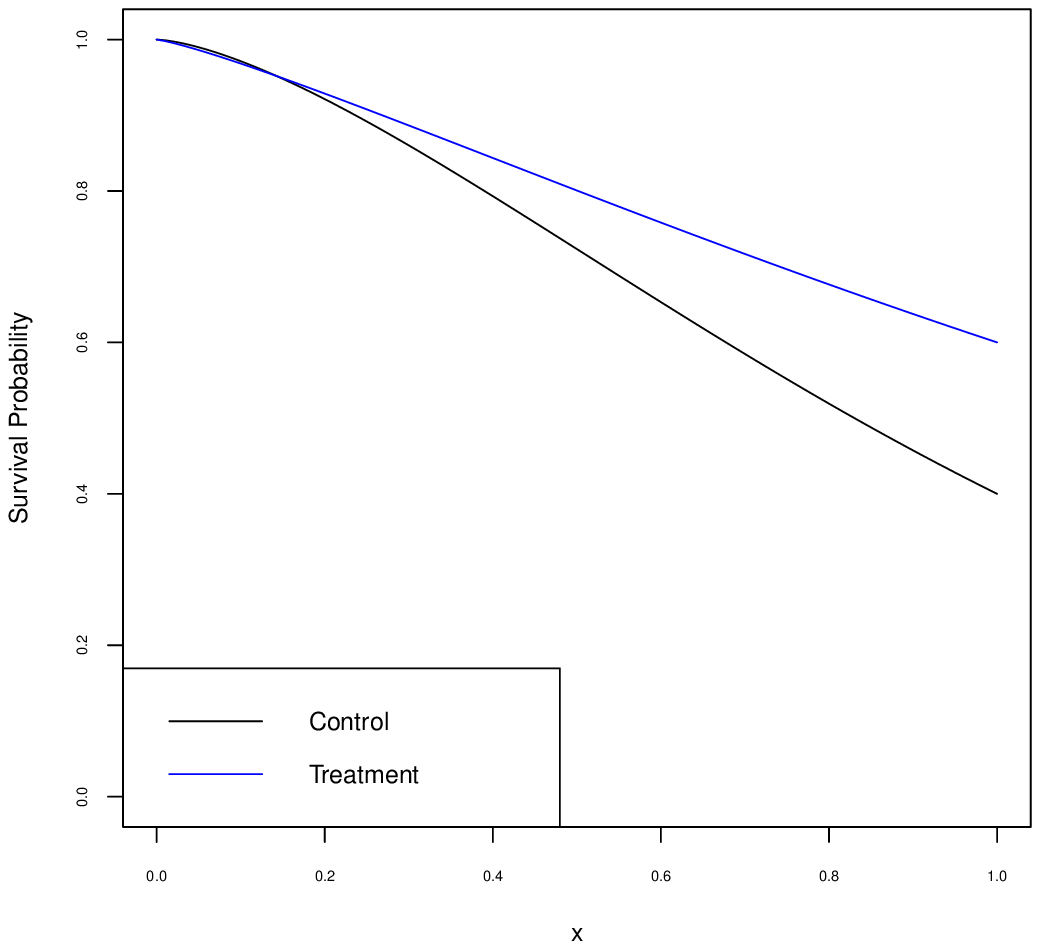}
 
  \vfill
 \label{fig:delayed}
\end{figure}

\begin{figure}[h]
 \caption{Cumulative Type I Error Rate and Power, PH Setting. Dashed lines indicate nominal stagewise type I error rate and power values under asymptotic distribution.}
  \centering
  \includegraphics[width=8cm]{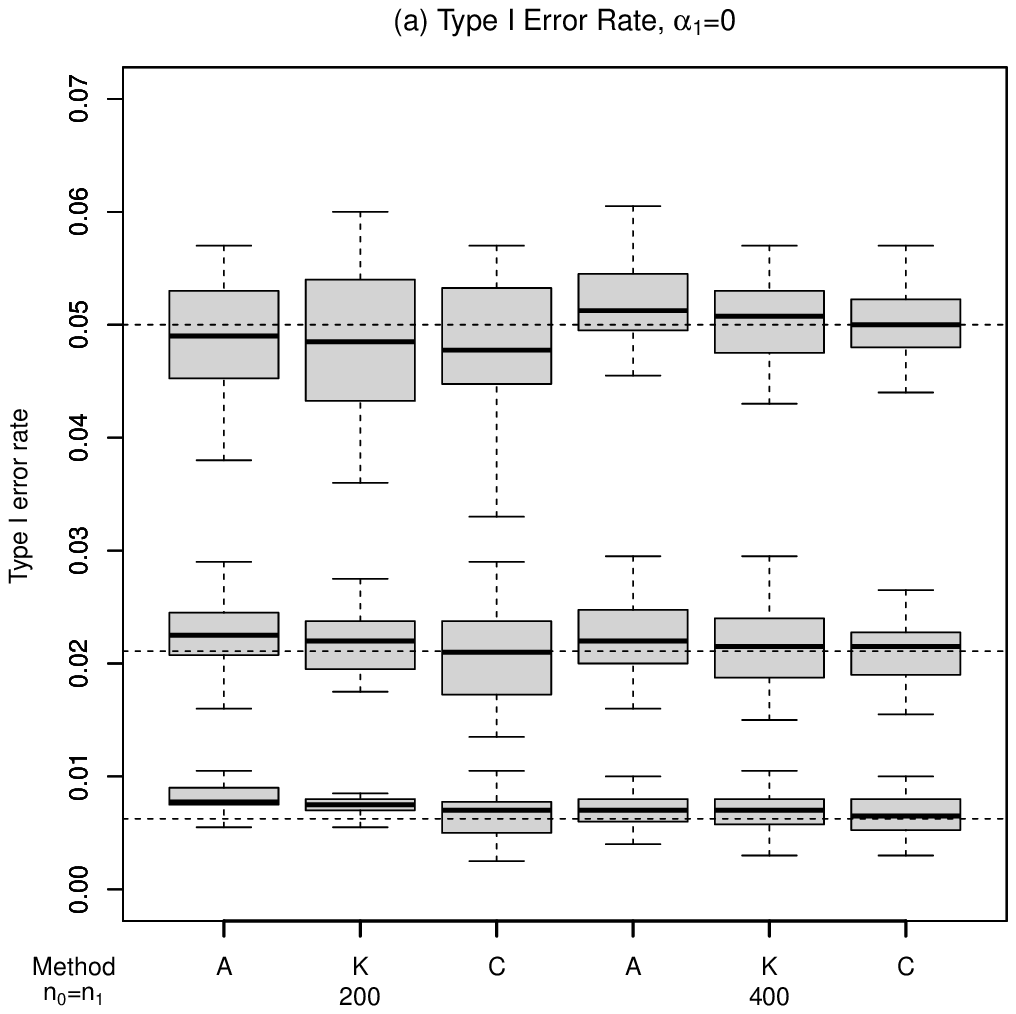}
  \includegraphics[width=8cm]{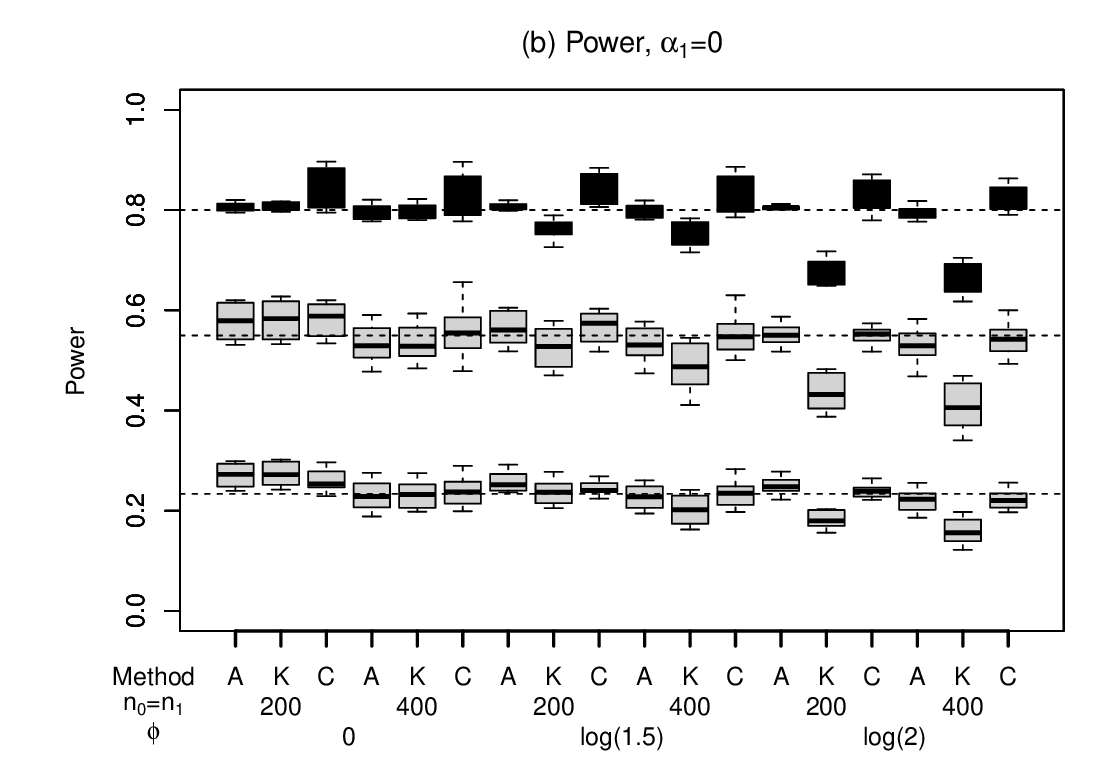}
  \vfill
  { \footnotesize A - Adjusted RMST, 
  K - Kaplan-Meier based RMST, C - Cox Model}
\label{fig:3.2}
\end{figure}

For each of the three methods, estimates of the cumulative type I error rate and power at each of the three stages were obtained.
The achieved stage type I error rates and power for the proposed method, 
Kaplan-Meier based RMST test, and Cox model based test under the NPH setting $\alpha_1 = -0.3$, are depicted in Figures \ref{fig:3.1}a and \ref{fig:3.1}b; this setting mimics a delayed treatment effect situation where the survival curves begin to separate at approximately $0.2 \tau$ years (Figure \ref{fig:delayed}) when an effect exists. 
Figure \ref{fig:3.1}a indicates that stagewise empirical type I error rates of both the proposed method and the Kaplan-Meier based RMST analysis attained the targeted levels, in particular achieving type I error rates of $5\%$ at the final stage. 
In contrast, the Cox model produced type I error rates of at least 10\% for all simulation settings in the NPH setting and does not adequately
control the type I error at a targeted 5\%. In the NPH setting, the type I error rate for the Cox model is increased because the null hypothesis of no difference in restricted mean survival times over the time horizon $[0, \tau]$ is not equivalent to the null hypothesis that the (average) hazard ratio is equal to 1. For example, when the parameters from the Weibull model are set to $\alpha_0=1.5, \alpha_1=-0.3, \gamma_0=-\log(0.4)$ to simulate a delayed effect setting (Figure \ref{fig:delayed}), with $\beta=\log(1.5)$, censoring rate set to 5\%, and uniform enrollment between $[0,2]$, the average hazard ratio is 0.94 under the null hypothesis of $\mu_0(1) = \mu_1(1)$. The Cox model tests the null hypothesis $\beta_W=0$, which is not true in the NPH simulation setting and leads to the increased type I error rate for the Cox model. 
Figure \ref{fig:3.1}b illustrates that the stagewise empirical power levels of the proposed method were 5-10\% higher than those of the Kaplan-Meier based analysis for non-zero values of $\phi$, implying that
adjusting for covariate(s) increases power compared to
not adjusting. Furthermore, the proposed test is more robust to covariate effect size than the test using a Kaplan-Meier based RMST analysis. The Cox model-based analysis attains the highest power, but its inflated type I error rates make it an unsuitable analysis approach in this NPH setting.

The achieved stagewise type I error and power for the three methods under the PH setting  ($\alpha_1 = 0$) are shown in Figures \ref{fig:3.2}a and \ref{fig:3.2}b.
The empirical type I error rates and power for the proposed method 
and Kaplan-Meier RMST based analysis followed similar trends as in the NPH setting, with the proposed method attaining 5-10\% higher power when influential covariates are at play. 
In the PH setting, 
the Cox model based test achieved the targeted type I error rate of $5\%$ and also obtained a modest boost in power of approximately 5\% compared to the proposed method.

\section{Example}
\label{s:example2}
Allogeneic bone marrow or blood stem cell transplant can be curative 
in patients who experience relapse of hematologic malignancies or receive a poor prognosis.
In these patients, a human leukocyte antigen (HLA)-matched unrelated donor or sibling is preferred because HLA mismatches
between recipient and donor are associated with an elevated probability of graft-versus-host-disease (GVHD), non-relapse mortality, and inferior survival.
However, a suitable HLA-matched donor may not be available for some patients; unrelated donor umbilical cord blood 
and an HLA-mismatched relative are alternative donor options.
These options motivated the Blood and Marrow Transplant Clinical Trials Network (BMT CTN) trial 1101, which
was a randomized multicenter phase 3 trial that
compared double umbilical cord blood (UCB) and haploidentical (Haplo) related donor transplantation for the treatment of leukemia and lymphoma in adults. The primary endpoint was progression-free survival (PFS) and important secondary endpoints were overall survival (OS),
non-relapse mortality, and malignancy relapse/progression \citep{Fuchs2020}.
During the design stage of BMT CTN 1101, because investigators had concerns that there may be NPH between the treatment groups,
the analysis of PFS utilized a fixed time point comparison
of survival probabilities at 2 years post-randomization.
In addition, since the trial was anticipated to last for 6 years,
a GS design was used with 3 planned interim  analyses.
The results from BMT CTN 1101 demonstrate that
(1) there is no significant difference in PFS between UCB
and Haplo transplantation for leukemia or lymphoma
and (2) Haplo transplantation provides lower non-relapse mortality and higher OS compared to UCB transplantation.

We re-visit the 1101 trial data and re-analyze the data using the proposed RMST method and compare the results with the original analysis's results.
In this trial, a total of 368 patients were randomized at a 1:1 allocation ratio to the two treatment arms, UCB and Haplo.
In the original trial, Kaplan-Meier estimates, which do not adjust for covariates, were used to compare PFS 
at 2 years post-randomization.
The original trial lasted longer than initially planned, with
an actual duration of 8 years. 
The enrollment period was 6 years and patients were followed for 2 years post-randomization. 
A group sequential test of Kaplan-Meier survival estimates with six planned interim analyses from years 3 to 8 annually was employed to compare PFS between treatment groups and also permit the possibility of declaring efficacy and stopping the trial early. 
The original analysis found that PFS did not differ between treatment groups significantly at any interim analysis; the trial did not conclude early. 
 At the last analysis, 2 year PFS estimates were 41\% for the Haplo arm and 35\% for the UCB arm ($p$=0.41).
  A significant difference was found for OS at 2 years;
 2 year estimates were 46\% for the UCB arm and 57\% for the Haplo arm ($p$=0.04).

 Multiple covariates are known or suspected to impact risk of progression or mortality in the patient population of interest, including age, gender, race, ethnicity, primary disease of lymphoma or leukemia, Karnofsky performance score, disease risk index, hematopoietic cell transplantation comorbidity index, and patient cytomegalovirus status at transplant.
 A previous secondary analysis did not find sufficient evidence that the PH assumption was violated for these covariates.
 With this in mind, 
 we re-analyze the 1101 trial data using using both the proposed covariate adjusted group sequential RMST method and a group sequential Kaplan-Meier based RMST analysis to compare PFS and OS at 2 years post-randomization between the treatment arms Haplo and UCB. 
 To permit early stopping for efficacy, alpha spending functions were used and selected from the power family in \cite{JennTurn99} with
$\rho =  3$. Yearly interim analyses were considered, beginning at 3 years after study opening.
To calculate information fractions for both PFS and OS during the re-analysis, we used the observed information at each stage divided by the total observed information from the trial.
Because the achieved information fractions for both outcomes were greater than 99.5\% at year 7, the re-analysis was concluded at year 7. 
For PFS, the total information was set at 0.00139; the observed information at year 7. For OS the total information was set at 0.00164, the observed information at year 7. 
Since the information fractions are different between PFS and OS,
different critical values were obtained from the alpha spending function. 

\begin{figure}[h]
 \caption{Group Sequential Test Statistics and Critical Values, Progression-free Survival. Dashed lines indicate critical values; solid lines indicate test statistics. The proposed method found a significant difference at stage 3, while the Kaplan-Meier based RMST comparison found a significant difference at stage 5.}
  \centering
  \includegraphics[width=8cm]{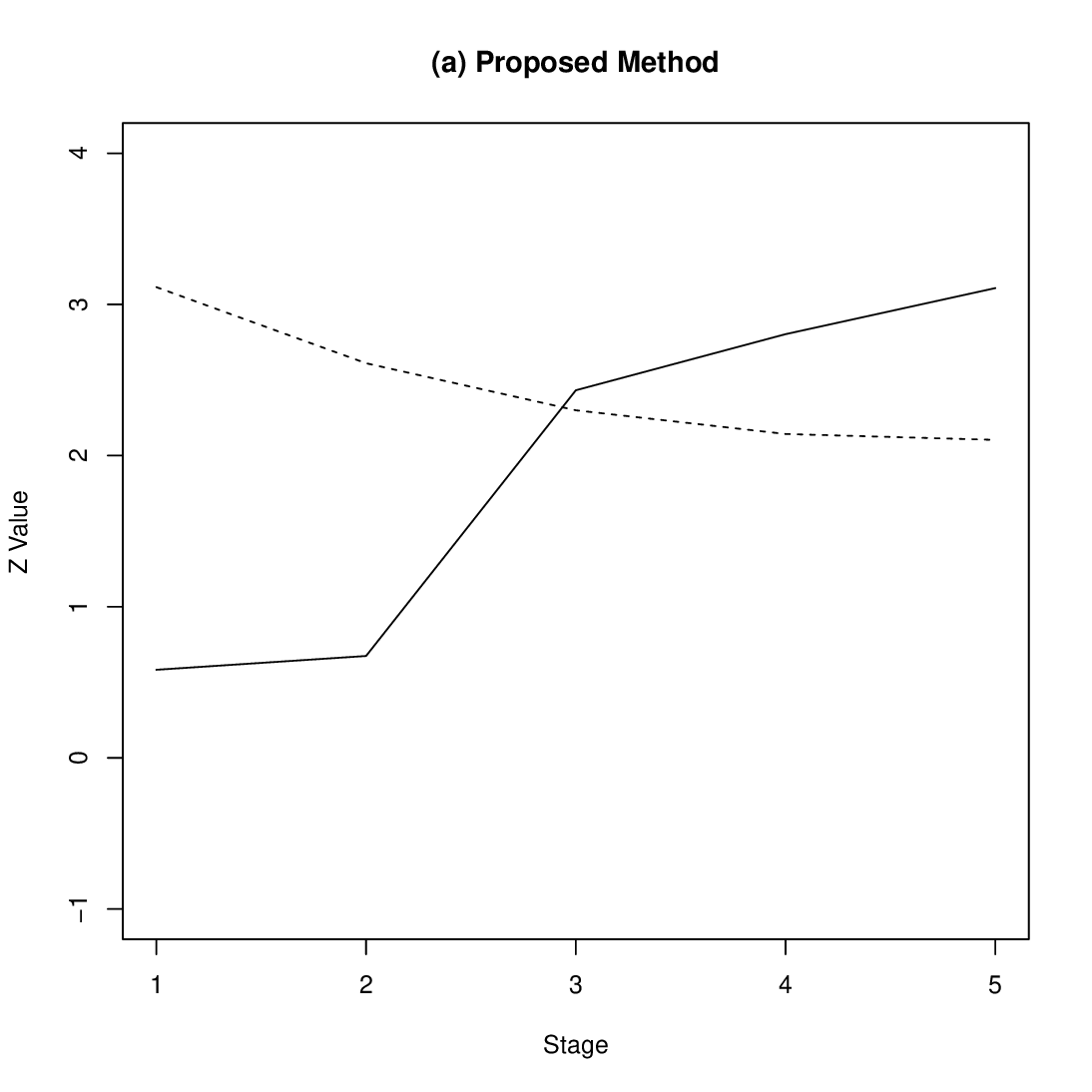}
  \includegraphics[width=8cm]{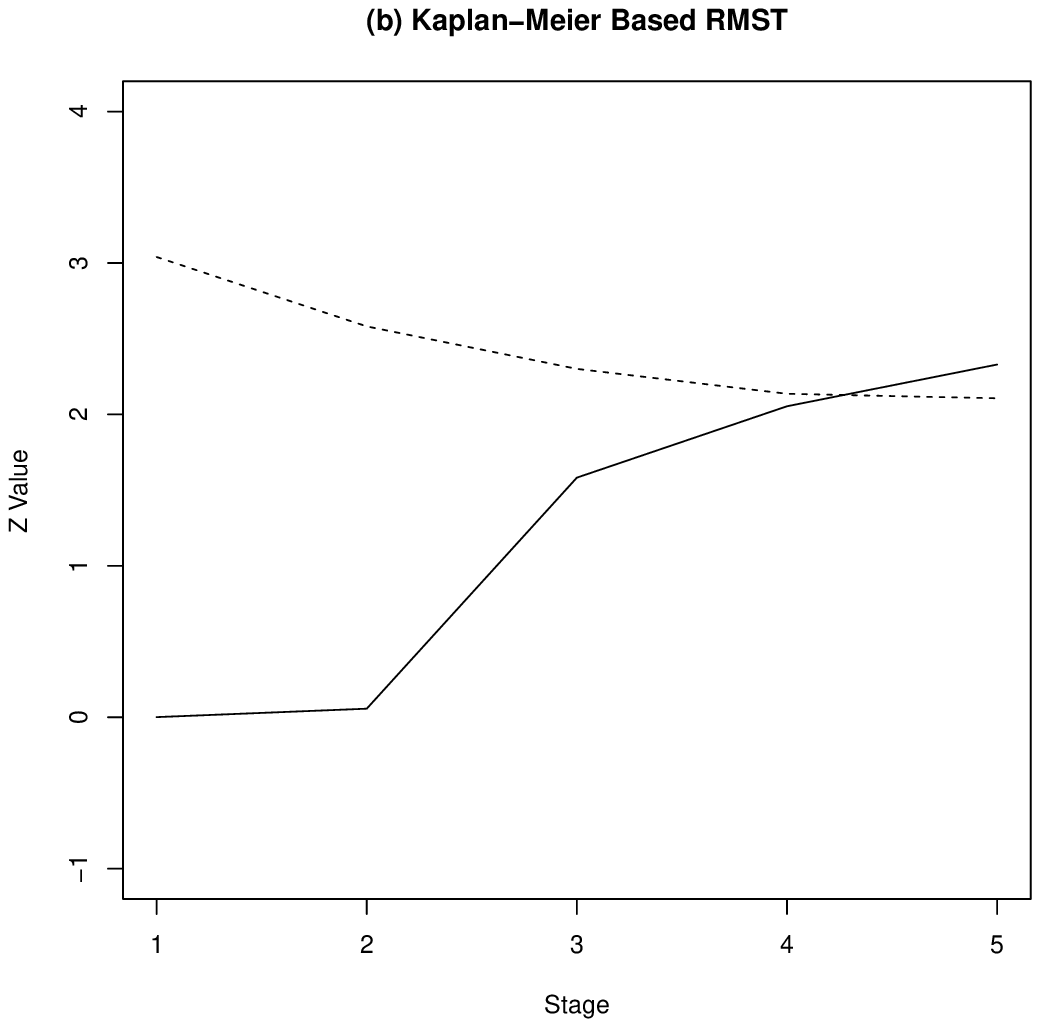}
  \vfill
\label{fig:4.1}
\end{figure}

Figures \ref{fig:4.1}a and \ref{fig:4.1}b show the standardized test statistics for the group sequential test using the proposed RMST method and the group sequential Kaplan-Meier based RMST test statistics, repsectively, for PFS. 
Both methods found significant differences and indicated higher 2 year PFS for the Haplo arm compared to UCB; the proposed method found a significant difference at stage 3 and the Kaplan-Meier RMST based method found a significant difference at stage 5. 
By adjusting for covariates and utilizing a covariate adjusted RMST based group sequential design, a significant difference in PFS
would be found 2 years before the Kaplan-Meier based RMST group sequential test found a significant difference.   

\begin{figure}[h]
 \caption{Group Sequential Test Statistics and Critical Values, Overall Survival. Dashed lines indicate critical values; solid lines indicate test statistics. Both the proposed method and Kaplan-Meier based RMST comparison found significant differences at stage 4.}
  \centering
  \includegraphics[width=8cm]{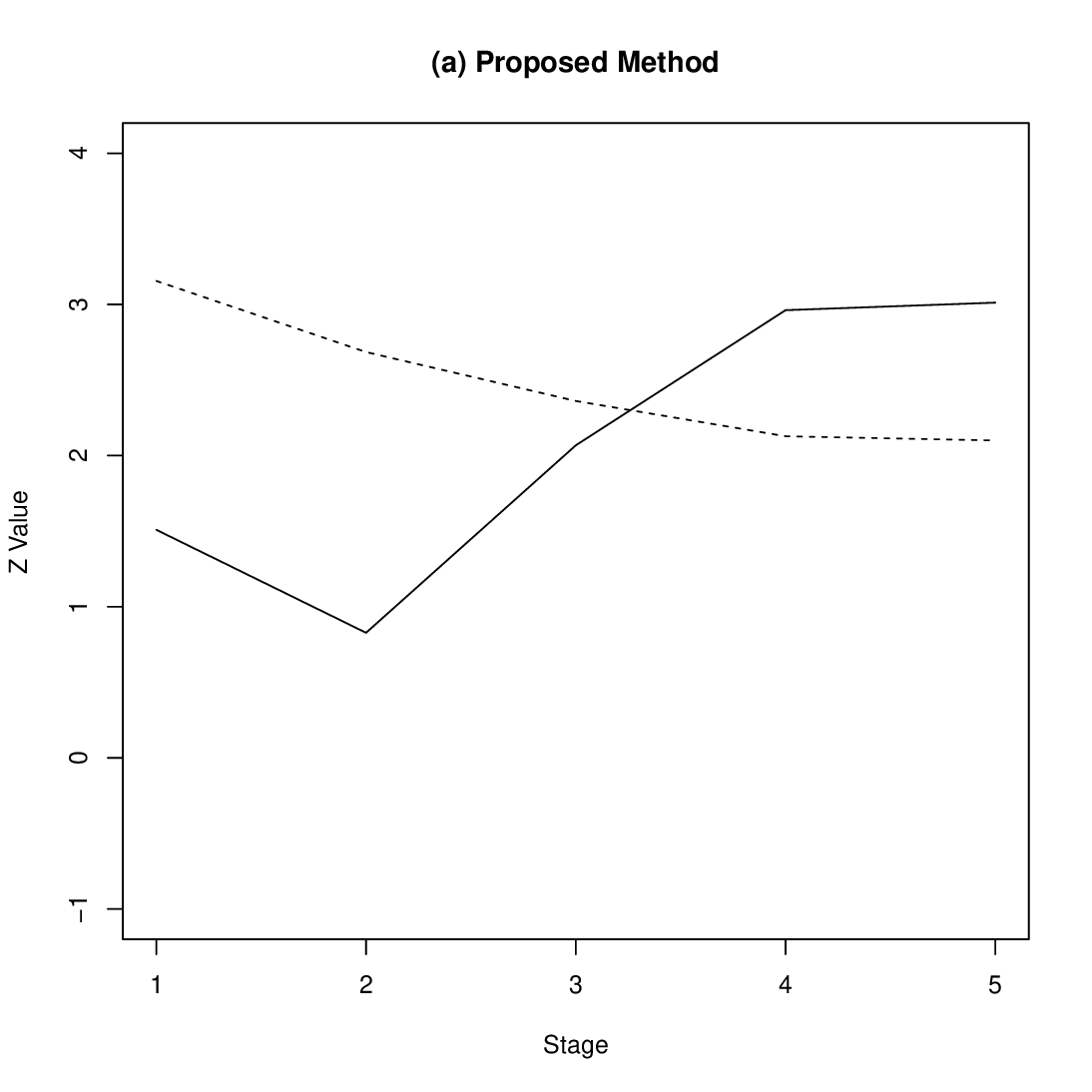}
  \includegraphics[width=8cm]{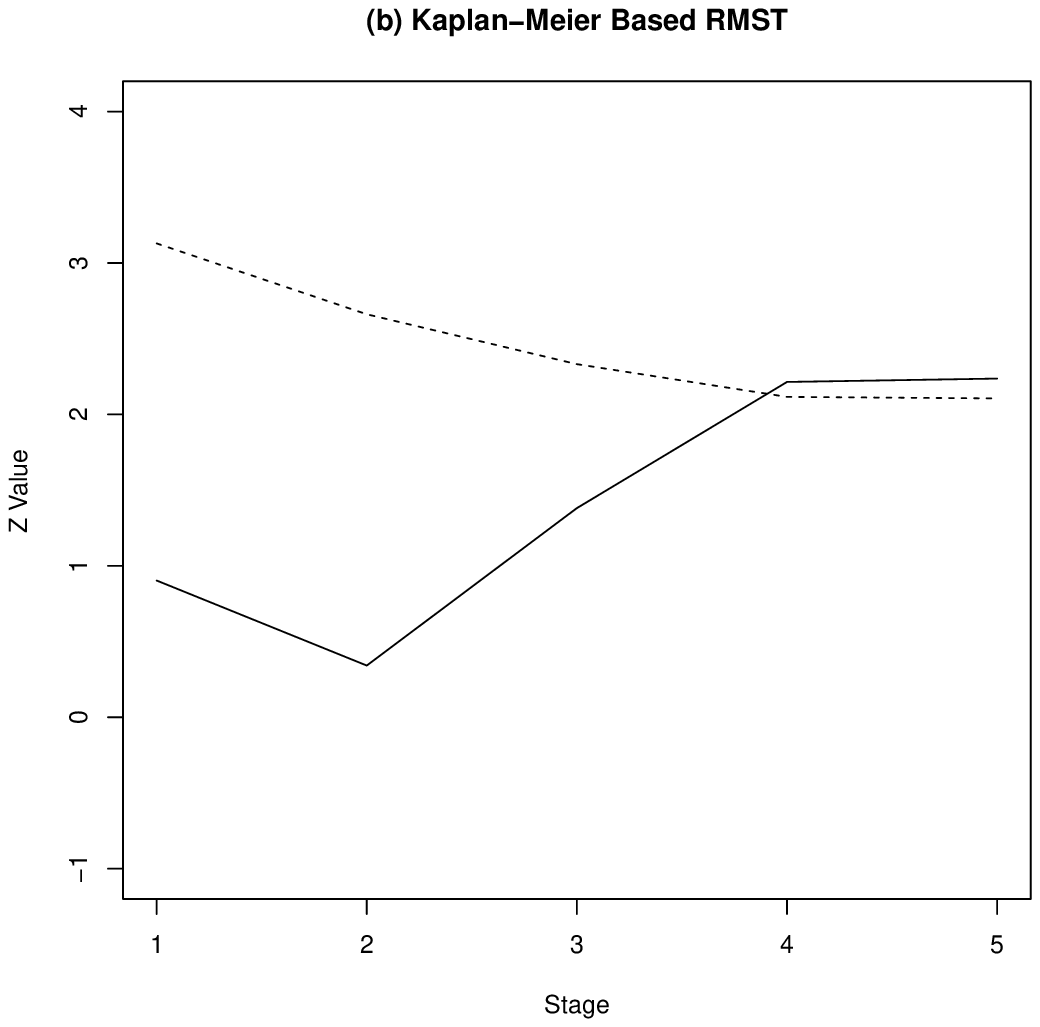}
  \vfill
\label{fig:4.2}
\end{figure}

Figures \ref{fig:4.2}a and \ref{fig:4.2}b show the standardized test statistics for the group sequential test using the proposed RMST method and the group sequential Kaplan-Meier based RMST test statistics, repsectively, for OS. 
Both methods found significant differences at stage 4 and indicated higher 2 year survival for the Haplo arm compared to UCB.

\section{Discussion}
In this paper, we have established a covariate-adjusted, RMST based group sequential test that avoids the PH assumption made by many existing methods.
We have derived the large sample distribution of the proposed adjusted group sequential test statistics and showed that they follow an independent increments structure,
which enables the use of existing methods to calculate critical values
for hypothesis testing. 
Our simulations indicate that the proposed group sequential tests meet targeted type I error rate and power specifications for realistic sample sizes.
We also found that the proposed method is robust to both the PH assumption and covariate effect size. 
The proposed methods were demonstrated with a re-analysis of the BMT CTN 1101 trial. 

Some extensions to the methods discussed in this paper are of interest.
The first extension is to sample size and power calculation methods to facilitate the design of clinical trials that will use the adjusted RMST as an endpoint. Future work will examine methods for accurate sample size and power calculations for fixed sample
and group sequential studies that use the adjusted RMST methods. 
 In this paper, type I error spending functions were employed to find critical values for GS testing for efficacy, so another extension is to consider type II error spending functions to permit GS testing for futility while attaining the desired type I error rate and power.
It is also of interest to examine GS testing of adjusted SPs and adjusted RMSTs in a clustered survival data setting, such as treatment center effects. A stratified marginal Cox model has been developed \citep{Lin94} that both accounts for correlated survival times within clusters and permits nonproportional hazards for the stratification variable. By stratifying the model on treatment arm, this model could potentially be employed in a GS analysis as well to compare arms under nonproportional treatment effect scenarios. 
Another possible extension is to investigate GS testing of the restricted mean time lost (RMTL) \citep{Ande13}, which is the competing risks analogue of RMST defined as the area under the cumulative incidence curve within a pre-specified interval time. The RMTL has a useful clinical interpretation as the mean amount of survival time lost due to a specific cause. 
Two methods for estimating the RMTLs of treatment groups are obtained by computing the areas under the Aalen-Johansen curve \citep{AaleJoha78} and the Fine-Gray model adjusted cumulative incidence curve \citep{ZhanZhan11}, the latter allowing covariate adjustment. Like the adjusted RMST estimator, these methods remain valid under nonproportional subdistribution hazard settings, and so are more robust than commonly used competing risks analysis methods such as Gray's test \citep{Gray88} and the Fine-Gray model \citep{FineGray99}. These RMTL analysis techniques have not been studied in the GS setting; we plan to investigate this extension in the future.

\section*{Supplementary Material}
Details on deriving Theorems 1 and 2 in Section 2 are available with this paper in the Appendix.

\section*{Acknowledgments}

The authors are grateful to the Blood and Marrow Transplant Clinical Trials Network (BMT CTN) for permission to use and re-analyze the 1101 trial data. Support for the BMT CTN 1101 trial was provided by grants U10HL069294 and U24HL138660 to the BMT CTN from the National Heart, Lung, and Blood Institute (NHLBI) and the National Cancer Institute (NCI). The Center for International Blood and Marrow Transplant Research (CIBMTR) is supported primarily by NCI, NHLBI, and the National Institute of Allergy and Infectious Diseases grant 5U24CA076518; and from HHSH234200637015C (HRSA/DHHS) to the Center for International Blood and Marrow Transplant Research; NHLBI and NCI grant 4U10HL069294; the Health Resources and Services Administration, Department of Health and Human Services contract HHSH250201200016C; and the Office of Naval Research grants N00014-17-1-2388 and N00014-16-1-2020. The content is solely the responsibility of the authors and does not necessarily represent the official views of the aformentioned parties.

\section*{Data Availability Statement}
The data used for the examples in Section 4 are available
from the Blood and Marrow Transplant Clinical Trials Network. Restrictions apply to the availability
of these data, which were permitted to be analyzed by the
authors in this paper. Data are available from the authors with the permission of the Blood and Marrow Transplant Clinical Trials Network.


\section*{Appendix: Proofs of Theorems 1 and 2}

\section{Regularity conditions}
\label{sec:3_6}
Throughout this section, assume that model \ref{eq:stratcox2} holds and
let $\bfbeta_0$ denote the true value of $\bfbeta$. 
For a vector $\bf y$, write $||\bfy|| = \max_i |({\bfy})_i|$ and $|{\bf y}| = \sqrt {\sum y_i^2}$.
For a matrix $\bfY$, write $||{\bfY}|| = \max_{i, j} |({\bfY})_{i, j}|$.
Throughout this section,
let $\bfbeta_0$ denote the true value of $\bfbeta$ and assume the following regularity conditions under model \ref{eq:stratcox2}. Define $\mathcal{D} = \{(u,t): 0 \le t \le u \le u^*, 0 \le t \le \tau\}$, where $u^*$ is an upper limit on the calendar age of the study.
The following regularity conditions are
assumed, which are similar to those of \cite{Flem11} under the unstratified proportional hazards regression model.

\bd
\item
(A) The survival time $\tau$ satisfies $\Lambda_{0i}(\tau) < \infty$ for $i=0,1$.

\item
(B) 
There exists a neighborhood ${\cal B}$ of $\bfbeta_0$ and, respectively, scalar, vector, and matrix functions
$s_{0i}(\bfbeta,u,t)$, $\mathbf{s}_{1i}(\bfbeta,u,t)$, and $\mathbf{s}_{2i}(\bfbeta,u,t)$
defined on ${\cal B} \times \mathcal{D}$ such that \\
$\sup_{(u,t) \in \mathcal{D}, \bfbeta \in {\cal B}} \bigl|\bigl|\mathbf{S}_{ki}(\bfbeta,u,t)-\mathbf{s}_{ki}(\bfbeta,u,t)\bigl|\bigl|
\parrow 0$ as $n \to \infty$ for $i = 0, 1$ and $k=0, 1, 2$.
Write
$$\mathbf{e}_i(\bfbeta,u,t) = {\mathbf{s}_{1i}(\bfbeta,u,t) \over s_{0i}(\bfbeta,u,t)}, \qquad
\mathbf{v}_i(\bfbeta,u,t) = {\mathbf{s}_{2i}(\bfbeta,u,t) \over s_{0i}(\bfbeta,u,t)} - \mathbf{e}_i(\bfbeta,u,t) \mathbf{e}_i^T(\bfbeta,u,t).$$
Due to the independence of $E_{ij}$ and $(T_{ij}, C_{ij}, \bfZ_{ij})$, 
the quantities $\mathbf{e}_i(\bfbeta,u,t)$ and $\mathbf{v}_i(\bfbeta,u,t)$ do not involve the calendar time $u$ under model \ref{eq:stratcox2}
and hence we can simply write $\mathbf{e}_i(\bfbeta,u,t)=\mathbf{e}_i(\bfbeta,t)$ and $\mathbf{v}_i(\bfbeta,u,t)=\mathbf{v}_i(\bfbeta,t)$ for $i=0,1$.

\item
(C) There exists a $\delta > 0$ such that, for $i = 0, 1$,
$${1 \over \sqrt{n_i}} \sup_{1 \leq j \leq n_i, (u,t) \in \mathcal{D}} \bigl|\bfZ_{ij}\bigl| Y_{ij}(u,t)
I(\bfbeta_0^T \bfZ_{ij} > - \delta |\bfZ_{ij}|) \parrow 0, \qquad {\rm as}~n \to \infty.$$

\item
(D) For $i=0,1$ and $k = 0, 1, 2$,
the functions $\mathbf{s}_{ki}(\bfbeta,u,t)$ are bounded and $s_{0i}(\bfbeta,u,t)$ is bounded away from 0 on
${\cal B} \times \mathcal{D}$.
The family of functions $\mathbf{s}_{ki}(\cdot,u,t)$, $(u,t) \in \mathcal{D}$, is an equicontinuous family at $\bfbeta_0$.

\item
(E) There exists constants $\rho_i \in (0, 1)$ such that
${n_i \over n} \longrightarrow \rho_i$ as $n \to \infty$ for each $i = 0,1$.

\item
(F) There exists $u_* \in [0,u^*]$ such that
the matrix $\boldsymbol{\Sigma}(\bfbeta_0,u,t)$ is positive definite for all $(u,t) \in \mathcal{D}_*$, where $\mathcal{D}_* = \{(u,t): u_* \le t \le u \le u^*, u_* \le t \le \tau\}$ and
\Eq{align*}
{
\boldsymbol{\Sigma}(\bfbeta,u,t)
=\sum_{i=0}^1 \rho_i \int_0^t \mathbf{v}_i(\bfbeta,u,s)s_{0i}(\bfbeta,u,s) \lambda_{0i}(s)ds.
}
Write $\boldsymbol{\Sigma}(\bfbeta,u)=\boldsymbol{\Sigma}(\bfbeta,u,u)$.

\item
(G) For each $i = 0,1$, there exists a function $\pi_i(u,t)$ with $\pi_i(u^*,\tau)>0$ such that
$$\sup_{(u,t) \in \mathcal{D}} \biggl|{1 \over n_i}\sum_{j=1}^{n_i}Y_{ij}(u,t) - \pi_i(u,t)\biggl| \parrow 0.$$
as $n \to \infty$.
\ed

In addition, we will use the following notations in the proofs of Theorems 1 and 2. For $i=0,1$, write
\Eq{align}
{
& C_i(\bfbeta,u,t)
= {\sqrt{n} \over n_i} \sum_{j=1}^{n_i}\int_0^t {1 \over S_{0i}(\bfbeta,u,s)}
I\biggl(\sum_{j=1}^{n_i}Y_{ij}(u,s) > 0\biggr) M_{ij}(u,ds), \cr
& \gamma_i(\bfbeta,u,t) = \int_0^t {1 \over s_{0i}(\bfbeta,u,s)} d\Lambda_{0i}(s), \qquad
\mathbf{Q}_i(\bfbeta,t) = \int_0^t \mathbf{e}_i(\bfbeta,s) d\Lambda_{0i}(s), \cr
& c_{i1}(t) = \sum_{g=0}^1 \rho_g E\{S_i(t|\bfZ_{g1})\exp(\bfbeta_0^T\bfZ_{g1})\}, \quad
\mathbf{c}_{i2}(t) = \sum_{g=0}^1 \rho_g E\{S_i(t|\bfZ_{g1})\exp(\bfbeta_0^T\bfZ_{g1})\bfZ_{g1}\}, \cr
& \mathbf{D}_i(\bfbeta,t) = c_{i1}(t) \mathbf{Q}_i(\bfbeta,t) - \Lambda_{0i}(t) \mathbf{c}_{i2}(t), \quad
\mathbf{D}(\bfbeta,t) = \mathbf{D}_1(\bfbeta,t) - \mathbf{D}_0(\bfbeta,t). \hskip0.4in
}

In order to make use of all available data at calendar time $u$ to estimate $\boldsymbol{\beta}$ from model \ref{eq:stratcox2}, the estimator $\hat{\boldsymbol{\beta}}(u,\min\{u,\tau\})$ should be used. For the remainder of this section, we assume this is done for all $u$ according to $\hat{\boldsymbol{\beta}}(u) = \hat{\boldsymbol{\beta}}(u,\min\{u,\tau\})$, suppressing the second argument. We first establish the asymptotic distributions of the two-parameter processes \\
$\bigl\{\sqrt{n} \bigl\{\hat S_1\bigl(\hat\bfbeta(u),u,t\bigr)- \hat S_0\bigl(\hat\bfbeta(u),u,t\bigr)\bigr\}
-\sqrt{n}\{\tilde S_1(t)-\tilde S_0(t)\}:~ (u,t) \in \mathcal{D} \bigr\}$
and \\ $\bigl\{\sqrt{n} \bigl\{\hat S_1\bigl(\hat\bfbeta(u),u,t\bigr)- \hat S_0\bigl(\hat\bfbeta(u),u,t\bigr)\bigr\}
-\sqrt{n}\{S_1(t)-S_0(t)\}:~ (u,t) \in \mathcal{D} \bigr\}$,
which are needed in proving Theorems 1 and 2.

\section{Asymptotic Distribution of Processes From the Treatment- Stratified Proportional Hazards Model With Staggered Entry}
\label{s:Processes}

By extending the work of \cite{Bila97}
from the unstratified proportional hazards regression model with staggered entry to
the treatment-stratified proportional hazards regression model with staggered entry,
we can establish the asymptotic distributions of the aforementioned two-parameter stochastic processes
under the regularity conditions (A)-(H) in Section \ref{sec:3_6}.
Under model \ref{eq:stratcox2}, since the treatment and control samples are independent,
it can be shown that the two-parameter score process
$\bigl\{n^{-1/2}\mathbf{U}(\bfbeta_0,u,t),~ (u,t) \in \mathcal{D}\bigr\}$
converges in distribution to a vector-valued Gaussian process
$\boldsymbol{\zeta}_1$
that has continuous sample paths, mean $\mathbf{0}$, and covariance function specified by
\Eq{align}
{
\label{eq:Ucovar}
& K_{\boldsymbol{\zeta}_1}\bigl((u_1,t_1),(u_2,t_2)\bigr) = E\{\boldsymbol{\zeta}_1(u_1,t_1)\boldsymbol{\zeta}_1^T(u_2,t_2)\}
= \boldsymbol{\Sigma}(\bfbeta_0,u_1 \wedge u_2,t_1 \wedge t_2),
}
for $(u_1,t_1), (u_2,t_2) \in \mathcal{D}$,
where $\boldsymbol{\Sigma}(\bfbeta,u,t)$ is defined in Condition (F) from Section {\ref{sec:3_6}.

As with Theorems 4.1 and 4.2 of \cite{Bila97},
we can show that $\hat\bfbeta(u,t)$ is strongly consistent for $\bfbeta_0$
uniformly for $(u,t) \in \mathcal{D}_* = \{(u,t): u_* \le t \le u \le u^*, t \le \tau\}$.
Then, a first-order Taylor series expansion gives
\Eq{align}
{
\label{eq:betaTaylor}
\sqrt{n}\bigl\{\hat\bfbeta(u,t) - \bfbeta_0\bigr\}
= \boldsymbol{\Sigma}^{-1}(\bfbeta_0,u,t) {1 \over \sqrt n} \boldsymbol{U}(\bfbeta_0,u,t) + o_p(1)
}
as $n \to \infty$, where $o_p(1)$ is uniform on $\mathcal{D}_*$.
This, along with \ref{eq:Ucovar}, implies that the two-parameter process
$\bigl\{\sqrt{n}\{\hat\bfbeta(u,t)-\bfbeta_0\},~ (u,t) \in \mathcal{D}_* \bigr\}$
converges in distribution to a vector-valued Gaussian random process
$\boldsymbol{\zeta}_2$
that has continuous sample path, mean $\mathbf{0}$, and covariance function specified by
\Eq{align}
{
\label{eq:betacovar}
& \mathbf{K}_{\boldsymbol{\zeta}_2}\bigl((u_1,t_1),(u_2,t_2)\bigr) = E\{\boldsymbol{\zeta}_2(u_1,t_1)\boldsymbol{\zeta}_2^{\tau}(u_2,t_2)\} \cr
& = \boldsymbol{\Sigma}^{-1}(\bfbeta_0,u_1,t_1) \boldsymbol{\Sigma}(\bfbeta_0,u_1 \wedge u_2, t_1 \wedge t_2) \boldsymbol{\Sigma}^{-1}(\bfbeta_0,u_2,t_2)
}
for $(u_1,t_1), (u_2,t_2) \in \mathcal{D}_*$.

For the treatment-specific baseline cumulative hazard function estimator \\ $\hat\Lambda_{0i}\bigl(\hat\bfbeta(u,t),u,t)$,
a first-order Taylor expansion yields, as $n \to \infty$,
\Eq{align}
{
\label{eq:lamdaTaylor}
& \sqrt{n} \bigl\{\hat\Lambda_{0i}\bigl(\hat\bfbeta(u,t),u,t\bigr) - \Lambda_{0i}(t)\bigr\} \cr
& = \sqrt{n} \bigl\{\hat\Lambda_{0i}(\bfbeta_0;u,t) - \Lambda_{0i}(t)\bigr\}
- \mathbf{Q}_i^T\bigl(\bfbeta_0,t\bigr) \boldsymbol{\Sigma}^{-1}(\bfbeta_0,u) {1 \over \sqrt{n}} \mathbf{U}(\bfbeta_0,u) \cr
& \quad + o_p(1) 
}
for $i=0,1$,
where $o_p(1)$ is uniform on $\mathcal{D}_*$ and
$\mathbf{Q}_i(\bfbeta,t)$ is defined in Section \ref{sec:3_6}.
Similar to the proof of Theorem 4.3 of \cite{Bila97}, the two-parameter process
$\bigl\{\sqrt{n}\{\hat\Lambda_{0i}\bigl(\bfbeta_0,u,t)-\Lambda_{0i}(t)\},~ (u,t) \in \mathcal{D}_* \bigr\}$
can be shown to be tight.
Furthermore, we can write
$\sqrt{n}\{\hat\Lambda_{0i}\bigl(\bfbeta_0,u,t\bigr) - \Lambda_{0i}(t)\} = {C}_i(\bfbeta_0,u,t)+o_p(1)$
as $n \to \infty$, where
${C}_i(\bfbeta,u,t)$ is defined in Section \ref{sec:3_6}.
This implies that the process
$\bigl\{\sqrt{n}\{\hat\Lambda_{0i}(\bfbeta_0,u,t) - \Lambda_{0i}(t)\},~ (u,t) \in \mathcal{D}_* \bigr\}$
converges in distribution to a Gaussian processes
$\zeta_{3i}$
with mean 0 and covariance function
\Eq{align}
{
\label{eq:lambdacovar}
K_{\zeta_{3i}}\bigl((u_1,t_1),(u_2,t_2)\bigr)
= E\{\zeta_{3i}(u_1,t_1)\zeta_{3i}(u_2,t_2)\}
= {1 \over \rho_i} \gamma_i(\bfbeta_0,u_1 \vee u_2, t_1 \wedge t_2\bigr)
}
for $(u_1,t_1), (u_2,t_2) \in \mathcal{D}_*$ and $i=0,1$, 
where $\gamma_i(\bfbeta,u,t)$ is defined in Section \ref{sec:3_6}.
Moreover, the processes $\zeta_{30}$, $\zeta_{31}$, and $(\zeta_1,\zeta_2)$ are independent.
These results, together with the asymptotic expansion in \ref{eq:lamdaTaylor},
imply that the two-parameter process
$\bigl\{\sqrt{n}\{\hat\Lambda_{0i}\bigl(\hat\bfbeta(u),u,t)-\Lambda_{0i}(t)\},~ u,t) \in \mathcal{D}_* \bigr\}$
converges in distribution to a Gaussian random process $\zeta_{4i}$ with mean 0 and covariance function specified by
\Eq{align*}
{
& K_{\zeta_{4i}}\bigl((u_1,t_1),(u_2,t_2)\bigr) = E\{\zeta_{4i}(u_1,t_1)\zeta_{4i}(u_2,t_2)\} \cr
&\quad= {1 \over \rho_i} \gamma_i(\bfbeta_0,u_1 \vee u_2, t_1 \wedge t_2\bigr)
+ Q_i^T\bigl(\bfbeta_0,t_1\bigr) \Sigma^{-1}(\bfbeta_0,u_1 \vee u_2) Q_i\bigl(\bfbeta_0,t_2\bigr)
}
for $(u_1,t_1), (u_2,t_2) \in \mathcal{D}_*$ and $i=0,1$.

\section{ Asymptotic distribution of
\\ $\bigl\{\sqrt{n} \bigl\{\hat S_1\bigl(\hat\bfbeta(u),u,t\bigr)- \hat S_0\bigl(\hat\bfbeta(u),u,t\bigr)\bigr\}
-\sqrt{n}\{\tilde S_1(t)-\tilde S_0(t)\}\bigr\}$}
Using a first-order Taylor expansion gives

\Eq{align*}
{
& \hat S_i\bigl(\hat\bfbeta(u),u,t\bigr)
= {1 \over n} \sum_{g=0}^1\sum_{j=1}^{n_g}
\exp[-\exp\{\hat\bfbeta^T(u)\bfZ_{gj}\} \hat\Lambda_{0i}(\hat\bfbeta(u);u,t)] \cr
&= {1 \over n} \sum_{g=0}^1\sum_{j=1}^{n_g} \exp\{-\exp(\bfbeta_0^T\bfZ_{gj}) \Lambda_{0i}(t)\} \cr
& - {1 \over n} \sum_{g=0}^1\sum_{j=1}^{n_g}
\exp\{-\exp(\bfbeta_0^T\bfZ_{gj}) \Lambda_{0i}(t)\} \exp(\bfbeta_0^T\bfZ_{gj}) \bfZ_{gj}^T\Lambda_{0i}(t)
\{\hat\bfbeta(u)-\bfbeta_0\} \cr
& - {1 \over n} \sum_{g=0}^1\sum_{j=1}^{n_g}
\exp\{-\exp(\bfbeta_0^T\bfZ_{gj}) \Lambda_{0i}(t)\} \exp(\bfbeta_0^T\bfZ_{gj})
\{\hat\Lambda_{0i}(\hat\bfbeta(u);u,t)-\Lambda_{0i}(t)\} + o_p(n^{-1/2}) \cr
&= {1 \over n} \sum_{g=0}^1\sum_{j=1}^{n_g} \exp\{-\exp(\bfbeta_0^T\bfZ_{gj}) \Lambda_{0i}(t)\} \cr
& - {1 \over n} \sum_{g=0}^1\sum_{j=1}^{n_g} S_i(t|\bfZ_{gj}) \exp(\bfbeta_0^T\bfZ_{gj}) \bfZ_{gj}^T\Lambda_{0i}(t)
\{\hat\bfbeta(u)-\bfbeta_0\} \cr
& - {1 \over n} \sum_{g=0}^1\sum_{j=1}^{n_g} S_i(t|\bfZ_{gj}) \exp(\bfbeta_0^T\bfZ_{gj})
\{\hat\Lambda_{0i}(\hat\bfbeta(u);u,t)-\Lambda_{0i}(t)\} + o_p(n^{-1/2}) \cr
&= \tilde S_i(t) - \Lambda_{0i}(t)\mathbf{c}_{i2}^T(t) \{\hat\bfbeta(u)-\bfbeta_0\}
- c_{i1}(t) \{\hat\Lambda_{0i}(\hat\bfbeta(u),u,t)-\Lambda_{0i}(t)\} + o_p(n^{-1/2}),
}
where $o_p(n^{-1/2})$ is uniform on $\mathcal{D}$.
It follows from the discussion in Section \ref{s:Processes}
that the processes
$\bigl\{\sqrt{n}\{\hat\bfbeta(u)-\bfbeta_0\},~ u_* \leq u \leq u^* \bigr\}$ and
$\bigl\{\sqrt{n}\{\hat\Lambda_{0i}\bigl(\hat\bfbeta(u),u,t)-\Lambda_{0i}(t)\},~ (u,t) \in \mathcal{D}_* \bigr\}$
are tight for $i=0,1$.
This, together with
\Eq{align*}
{
& \sqrt{n} \bigl\{\hat S_i\bigl(\hat\bfbeta(u),u,t\bigr)-\tilde S_i(t)\bigr\} \cr
& = - \Lambda_{0i}(t)\mathbf{c}_{i2}^T(t) \sqrt{n}\{\hat\bfbeta(u)-\bfbeta_0\} \cr
& \quad - c_{i1}(t) \sqrt{n}\{\hat\Lambda_{0i}(\hat\bfbeta(u),u,t)-\Lambda_{0i}(t)\} + o_p(1),
}
implies that the process
$\bigl\{\sqrt{n} \bigl\{\hat S_i\bigl(\hat\bfbeta(u),u,t\bigr)-\tilde S_i(t)\bigr\}: 
~ (u,t) \in \mathcal{D}_* \bigr\}$ is tight.
Therefore, we conclude that the process
$\bigl\{\sqrt{n} \bigl\{\hat S_1\bigl(\hat\bfbeta(u),u,t\bigr)- \hat S_0\bigl(\hat\bfbeta(u),u,t\bigr)\bigr\}
-\sqrt{n}\{\tilde S_1(t)-\tilde S_0(t)\}:~ (u,t) \in \mathcal{D}_* \bigr\}$ is tight.
Now using the asymptotic expansions \ref{eq:betaTaylor} and \ref{eq:lamdaTaylor}, 
along with the asymptotic convergences of 
$\bigl\{\sqrt{n}\{\hat\bfbeta(u)-\bfbeta_0\}\bigr\}$ and
$\bigl\{\sqrt{n}\{\hat\Lambda_{0i}\bigl(\bfbeta_0,u,t)-\Lambda_{0i}(t)\}\bigr\}$
in \ref{eq:betacovar} and \ref{eq:lambdacovar} , we have, as $n \to \infty$,
\Eq{align}
{
\label{eq:expansion}
& \sqrt{n} \bigl\{\hat S_1\bigl(\hat\bfbeta(u),u,t\bigr)- \hat S_0\bigl(\hat\bfbeta(u),u,t\bigr)\bigr\}
-\sqrt{n}\{\tilde S_1(t)-\tilde S_0(t)\} \cr
&~~= \sqrt{n} \bigl\{\hat S_1\bigl(\hat\bfbeta(u),u,t\bigr)-\tilde S_1(t)\bigr\}
- \sqrt{n} \bigl\{\hat S_0\bigl(\hat\bfbeta(u),u,t\bigr)-\tilde S_0(t)\bigr\} \cr
&~~= \biggl[- \Lambda_{01}(t)\mathbf{c}_{12}^T(t) \sqrt{n}\{\hat\bfbeta(u)-\bfbeta_0\}
- c_{11}(t) \sqrt{n}\{\hat\Lambda_{01}(\hat\bfbeta(u),u,t)-\Lambda_{01}(t)\} + o_p(1) \biggr] \cr
&\qquad - \biggl[- \Lambda_{00}(t)\mathbf{c}_{02}^T(t) \sqrt{n}\{\hat\bfbeta(u)-\bfbeta_0\}
- c_{01}(t) \sqrt{n}\{\hat\Lambda_{00}(\hat\bfbeta(u),u,t)-\Lambda_{00}(t)\} + o_p(1) \biggr] \cr
&~~= \biggl[- c_{11}(t) C_1(\bfbeta_0,u,t)
+ \mathbf{D}_1^T(\bfbeta_0,t) \boldsymbol{\Sigma}^{-1}(\bfbeta_0,u) {1 \over \sqrt{n}} \mathbf{U}(\bfbeta_0,u) + o_p(1)\biggr] \cr
&\qquad -\biggl[- c_{01}(t) C_0(\bfbeta_0,u,t)
+ \mathbf{D}_0^T(\bfbeta_0,t) \boldsymbol{\Sigma}^{-1}(\bfbeta_0,u) {1 \over \sqrt{n}} \mathbf{U}(\bfbeta_0,u) + o_p(1)\biggr] \cr
&~~= c_{01}(t) C_0(\bfbeta_0,u,t)-c_{11}(t) C_1(\bfbeta_0,u,t) \cr
&\qquad +\{\mathbf{D}_1(\bfbeta_0,t)-\mathbf{D}_0(\bfbeta_0,t)\}^T
\boldsymbol{\Sigma}^{-1}(\bfbeta_0,u) {1 \over \sqrt{n}} \mathbf{U}(\bfbeta_0,u) + o_p(1) \cr
&~~\Darrow c_{01}(t) \zeta_{30}(u,t)-c_{11}(t) \zeta_{31}(u,t)
+\{\mathbf{D}_1(\bfbeta_0,t)-\mathbf{D}_0(\bfbeta_0,t)\}^T \boldsymbol{\zeta}_2(u,u).
}
Let 
$\zeta_5(u,t) 
= c_{01}(t) \zeta_{30}(u,t)-c_{11}(t) \zeta_{31}(u,t) +\mathbf{D}^T(\bfbeta_0,u,t) \boldsymbol{\zeta}_2(u,u)$.
Since the processes $\zeta_{30}$, $\zeta_{31}$, and $(\boldsymbol{\zeta}_1^T,\boldsymbol{\zeta}_2^T)^T$ are independent,
it follows from \ref{eq:betacovar} and \ref{eq:lambdacovar} that the two-parameter process
$\bigl\{\sqrt{n} \bigl\{\hat S_1\bigl(\hat\bfbeta(u),u,t\bigr)- \hat S_0\bigl(\hat\bfbeta(u),u,t\bigr)\bigr\}
-\sqrt{n}\{\tilde S_1(t)-\tilde S_0(t)\}:~ (u,t) \in \mathcal{D}_* \bigr\}$
converges in distribution to the Gaussian random process $\zeta_5$
with mean 0 and covariance function
\Eq{align}
{
\label{eq:COVAR}
& K_{\zeta_5}\bigl((u_1,t_1), (u_2,t_2)\bigr) = E\{\zeta_5(u_1,t_1)\zeta_5(u_2,t_2)\} \cr
&= E\biggl[\{c_{01}(t_1) \zeta_{30}(u_1,t_1)-c_{11}(t_1) \zeta_{31}(u_1,t_1)
+ \mathbf{D}^T(\bfbeta_0,t_1) \boldsymbol{\zeta}_2(u_1,u_1)\} \cr
&\qquad \times \{c_{01}(t_2) \zeta_{30}(u_2,t_2)-c_{11}(t_2) \zeta_{31}(u_2,t_2)
+\mathbf{D}^T(\bfbeta_0,t_2) \boldsymbol{\zeta}_2(u_2,u_2)\}\biggr] \cr
&= c_{01}(t_1)c_{01}(t_2)E\{\zeta_{30}(u_1,t_1)\zeta_{30}(u_2,t_2)\}
+ c_{11}(t_1)c_{11}(t_2)E\{\zeta_{31}(u_1,t_1)\zeta_{31}(u_2,t_2)\} \cr
&\qquad + \mathbf{D}^T(\bfbeta_0,t_1)E\{\boldsymbol{\zeta}_2(u_1,u_1)\boldsymbol{\zeta}_2^T(u_2,u_2)\} \mathbf{D}(\bfbeta_0,t_2)\cr
&= c_{01}(t_1)c_{01}(t_2) {1 \over \rho_0} \gamma_0(\bfbeta_0,u_1 \vee u_2, t_1 \wedge t_2\bigr)
+ c_{11}(t_1)c_{11}(t_2) {1 \over \rho_1} \gamma_1(\bfbeta_0,u_1 \vee u_2, t_1 \wedge t_2\bigr) \cr
&\qquad + \mathbf{D}^T(\bfbeta_0,t_1)
\boldsymbol{\Sigma}^{-1}(\bfbeta_0,u_1,t_1) \boldsymbol{\Sigma}(\bfbeta_0,u_1 \wedge u_2, t_1 \wedge t_2) \boldsymbol{\Sigma}^{-1}(\bfbeta_0,u_2,t_2)
\mathbf{D}^T(\bfbeta_0,t_2) \cr
&= c_{01}(t_1)c_{01}(t_2) {1 \over \rho_0} \gamma_0(\bfbeta_0,u_1 \vee u_2, t_1 \wedge t_2\bigr)
+ c_{11}(t_1)c_{11}(t_2) {1 \over \rho_1} \gamma_1(\bfbeta_0,u_1 \vee u_2, t_1 \wedge t_2\bigr) \cr
&\qquad + \mathbf{D}^T(\bfbeta_0,t_1) \boldsymbol{\Sigma}^{-1}(\bfbeta_0,u_1 \vee u_2, u_1 \vee u_2) \mathbf{D}^T(\bfbeta_0,t_2) \cr
&= E\{\zeta_5(u_1 \vee u_2,t_1)\zeta_5(u_1 \vee u_2,t_2)\}
= K_{\zeta_5}\bigl((u_1 \vee u_2,t_1), (u_1 \vee u_2,t_2)\bigr)
}
for $(u_1,t_1), (u_2,t_2) \in \mathcal{D}_*$.

\noindent
\section{Asymptotic distribution of \\
$\bigl\{\sqrt{n} \bigl\{\hat S_1\bigl(\hat\bfbeta(u),u,t\bigr)- \hat S_0\bigl(\hat\bfbeta(u),u,t\bigr)\bigr\}
-\sqrt{n}\{S_1(t)-S_0(t)\}\bigr\}$}
\smallskip

\noindent
We can write
\Eq{align}
{
\label{eq:tilde}
& \sqrt{n} \bigl\{\hat S_1\bigl(\hat\bfbeta(u),u,t\bigr)- \hat S_0\bigl(\hat\bfbeta(u),u,t\bigr)\bigr\}
- \sqrt{n}\{S_1(t)-S_0(t)\} \cr
&\quad= \sqrt{n}\bigl\{\hat S_1\bigl(\hat\bfbeta(u),u,t\bigr)- \hat S_0\bigl(\hat\bfbeta(u),u,t\bigr)\bigr\}
-\sqrt{n}\{\tilde S_1(t)-\tilde S_0(t)\} \cr
&\qquad + \sqrt{n} \{\tilde S_1(t) - \tilde S_1(t)\} - \sqrt{n} \{S_1(t)-S_0(t)\}.
}
Since the process
$\bigl\{\sqrt{n} \bigl\{\hat S_1\bigl(\hat\bfbeta(u),u,t\bigr)- \hat S_0\bigl(\hat\bfbeta(u),u,t\bigr)\bigr\}
-\sqrt{n}\{\tilde S_1(t)-\tilde S_0(t)\}:~ (u,t) \in \mathcal{D}_* \bigr\}$ 
converges in distribution to the Gaussian process $\zeta_5$, it is tight.
The multivariate central limit theorem implies that the process 
$\bigl\{\sqrt{n}\{\tilde S_i(t)-S_i(t)\}: ~ u_* \le t \leq \tau\bigr\}$  
converges in distribution to a Gaussian process and is tight for $i=0,1$.
This further implies that the process
$\bigl\{\sqrt{n}\{\tilde S_1(t)-\tilde S_0(t)\} - \sqrt{n} \{S_1(t)-S_0(t)\}: ~ u_* \le t \leq \tau\bigr\}$ is tight.
As a result, it follows from \ref{eq:expansion} and \ref{eq:tilde} that
$\sqrt{n} \bigl\{\hat S_1\bigl(\hat\bfbeta(u),u,t\bigr)- \hat S_0\bigl(\hat\bfbeta(u),u,t\bigr)\bigr\}
-\sqrt{n}\{S_1(t)-S_0(t)\}$
is asymptotically a sum of independent zero-mean random variables with a remainder term of order $o_p(1)$.
The multivariate central limit theorem and the aforementioned tightness now entail that
the two-parameter process
$\bigl\{\sqrt{n} \bigl\{\hat S_1\bigl(\hat\bfbeta(u),u,t\bigr)- \hat S_0\bigl(\hat\bfbeta(u),u,t\bigr)\bigr\}
-\sqrt{n}\{S_1(t)-S_0(t)\}:~ (u,t) \in \mathcal{D}_* \bigr\}$
converges in distribution to a Gaussian process $\zeta_6$ with mean 0.
The covariance function of $\zeta_6$ is calculated as
\Eq{align}
{
\label{eq:COVAR6}
& K_{\zeta_6}\bigl((u_1,t_1), (u_2,t_2)\bigr) = E\{\zeta_6(u_1,t_1)\zeta_6(u_2,t_2)\} \cr
&= \Cov\biggl[c_{01}(t_1) C_0(\bfbeta_0,u_1,t_1)-c_{11}(t_1) C_1(\bfbeta_0,u_1,t_0)
+ \mathbf{D}^T(\bfbeta_0,t_1) \boldsymbol{\Sigma}^{-1}(\bfbeta_0,u_1) {1 \over \sqrt{n}} \mathbf{U}(\bfbeta_0,u_1) \cr
&\qquad + \sqrt{n}\{\tilde S_1(t_1)-\tilde S_0(t_1)\} - \sqrt{n}\{S_1(t_1)-S_0(t_1)\}, \cr
&\qquad c_{01}(t_0) C_0(\bfbeta_0,u_2,t_2) -c_{11}(t_2) C_1(\bfbeta_0,u_2,t_2)
+ \mathbf{D}^T(\bfbeta_0,t_2) \boldsymbol{\Sigma}^{-1}(\bfbeta_0,u_2) {1 \over \sqrt{n}} \mathbf{U}(\bfbeta_0,u_2) \cr
&\qquad + \sqrt{n}\{\tilde S_1(t_2)-\tilde S_0(t_2)\} - \sqrt{n}\{S_1(t_2)-S_0(t_2)\}\biggr] \cr
&= \Cov\biggl[c_{01}(t_1) C_0(\bfbeta_0,u_1,t_1)-c_{11}(t_1) C_1(\bfbeta_0,u_1,t_1)
+ \mathbf{D}^T(\bfbeta_0,t_1) \boldsymbol{\Sigma}^{-1}(\bfbeta_0,u_1) {1 \over \sqrt{n}} \mathbf{U}(\bfbeta_0,u_1), \cr
&\qquad c_{01}(t_2) C_0(\bfbeta_0,u_2,t_2)-c_{11}(t_2) C_1(\bfbeta_0,u_2,t_2)
+ \mathbf{D}^T(\bfbeta_0,t_2) \boldsymbol{\Sigma}^{-1}(\bfbeta_0,u_2) {1 \over \sqrt{n}} \mathbf{U}(\bfbeta_0,u_2)\biggr] \cr
&\qquad + \Cov\bigl(\sqrt{n}\{\tilde S_1(t_1)-\tilde S_0(t_1)\},~
\sqrt{n}\{\tilde S_1(t_2)-\tilde S_0(t_2)\}\bigr) \cr
&= K_{\zeta_5}\bigl((u_1,t_1), (u_2,t_2)\bigr) 
+ \Cov[\sqrt{n}\{\tilde S_1(t_1)-\tilde S_0(t_1)\},~ \sqrt{n}\{\tilde S_1(t_2)-\tilde S_0(t_2)\}] \cr
&= K_{\zeta_5}\bigl((u_1,t_1), (u_2,t_2)\bigr) 
+ \Cov\{S_1(t_1|\bfZ)-S_0(t_1|\bfZ),~ S_1(t_2|\bfZ)-S_0(t_2|\bfZ)\}
}
for $(u_1,t_1), (u_2,t_2) \in \mathcal{D}_*$.

\section{Proof of Theorem 1}
Using the asymptotic expansion in \ref{eq:expansion} yields
\Eq{align*}
{
& \sqrt{n} \{\hat\Delta(u,\tau)-\tilde\Delta(u,\tau)\}
= \int_0^{\tau} \sqrt{n}\bigl[\bigl\{\hat S_1\bigl(\hat\bfbeta(u);u,t\bigr)
- \hat S_0\bigl(\hat\bfbeta(u);u,t\bigr)\bigr\} - \{\tilde S_1(t)-\tilde S_0(t)\}\bigr]dt \cr
&~~= \int_0^{\tau} \biggl[c_{01}(t) C_0(\bfbeta_0,u,t)-c_{11}(t) C_1(\bfbeta_0,u,t) \cr
& \qquad +\{\mathbf{D}_1(\bfbeta_0,t)-\mathbf{D}_0(\bfbeta_0,t)\}^T \boldsymbol{\Sigma}^{-1}(\bfbeta_0,u) 
{1 \over \sqrt{n}} \mathbf{U}(\bfbeta_0,u)\biggr]dt+o_p(1).
}
Since $\sqrt{n} \bigl\{\hat S_1\bigl(\hat\bfbeta(u),u,t\bigr)- \hat S_0\bigl(\hat\bfbeta(u),u,t\bigr)\bigr\}
-\sqrt{n}\{\tilde S_1(t)-\tilde S_0(t)\} \darrow \zeta_5(u,t)$,
$$\sqrt{n} \{\hat\Delta(u,\tau)-\tilde\Delta(u,\tau)\} \darrow \int_0^{\tau} \zeta_5(u,t)dt.$$
This implies that the stochastic process
$\bigl\{\sqrt{n} \{\hat\Delta(u,\tau)-\tilde\Delta(u,\tau)\},~ u_* \leq u \leq u^* \bigr\}$
converges in distribution to a Gaussian process $\xi$ 
with mean 0, where $\xi(u,\tau) = \int_0^{\tau} \zeta_5(u,t)dt$.
Using the covariance function of $\zeta_5$ in \ref{eq:COVAR},
the covariance function of $\xi$ is calculated as
\Eq{align*}
{
& K_{\xi}(u_1,u_2,\tau) = \Cov\{\xi(u_1,\tau), \xi(u_2,\tau)\} 
= E\{\xi(u_1,\tau)\xi(u_2,\tau)\} \cr
&\quad= E\biggl\{\int_0^{\tau} \zeta_5(u_1,s)ds \int_0^{\tau} \zeta_5(u_2,t)dt\biggr\}
= E\biggl\{\int_0^{\tau}\int_0^{\tau} \zeta_5(u_1,s)\zeta_5(u_2,t)dsdt\biggr\} \cr
&\quad= \int_0^{\tau}\int_0^{\tau} E\{\zeta_5(u_1,s)\zeta_5(u_2,t)\}dsdt
= \int_0^{\tau}\int_0^{\tau} K_{\zeta_5}\bigl((u_1,s),(u_2,t)\bigr)dsdt \cr
&\quad= \int_0^{\tau}\int_0^{\tau}
\biggl\{{1 \over \rho_0} c_{01}(s)c_{01}(t) \gamma_0(\bfbeta_0,u_1 \vee u_2, s \wedge t\bigr)
+ {1 \over \rho_1} c_{11}(s)c_{11}(t) \gamma_1(\bfbeta_0,u_1 \vee u_2, s \wedge t\bigr) \cr
&\qquad + \mathbf{D}^T(\bfbeta_0,s) \boldsymbol{\Sigma}^{-1}(\bfbeta_0,u_1 \vee u_2) \mathbf{D}(\bfbeta_0,t) \biggr\} dsdt \cr
&\quad= {1 \over \rho_0} \int_0^{\tau}\int_0^{\tau}
c_{01}(s)c_{01}(t) \gamma_0(\bfbeta_0, u_1 \vee u_2, s \wedge t\bigr)dsdt \cr
&\qquad + {1 \over \rho_1} \int_0^{\tau}\int_0^{\tau} c_{11}(s)c_{11}(t) 
\gamma_1(\bfbeta_0,u_1 \vee u_2, s \wedge t\bigr)dsdt \cr
&\qquad + \biggl\{\int_0^{\tau}\mathbf{D}^T(\bfbeta_0,s)ds\biggr\}
\boldsymbol{\Sigma}^{-1}(\bfbeta_0,u_1 \vee u_2) \biggl\{\int_0^{\tau} \mathbf{D}(\bfbeta_0,t)dt\biggr\} \cr
&\quad= {1 \over \rho_0} \int_0^{\tau}\int_0^{\tau}
c_{01}(s)c_{01}(t) \gamma_0(\bfbeta_0,u_1 \vee u_2, s \wedge t\bigr)dsdt \cr
&\qquad + {1 \over \rho_1} \int_0^{\tau}\int_0^{\tau} c_{11}(s)c_{11}(t) 
\gamma_1(\bfbeta_0,u_1 \vee u_2, s \wedge t\bigr)dsdt \cr
&\qquad + \biggl\{\int_0^{\tau} \mathbf{D}(\bfbeta_0,s)ds\biggr\}^T
\boldsymbol{\Sigma}^{-1}(\bfbeta_0,u_1 \vee u_2) \biggl\{\int_0^{\tau} \mathbf{D}(\bfbeta_0,t) dt\biggr\} \cr
&\quad= K_{\xi}(u_1 \vee u_2, u_1 \vee u_2,\tau) 
= \Var\{\xi(u_1 \vee u_2,\tau)\} = V_{\xi}^2(u_1 \vee u_2, \tau).
}
for $u_* \leq u_1, u_2 \leq u^*$.

\section{Proof of Theorem 2}

Using \ref{eq:expansion} and \ref{eq:tilde}, we can write, as $n \to \infty$,
\Eq{align*}
{
& \sqrt{n} \{\hat\Delta(u,\tau)-\Delta(u,\tau)\} \\ 
& = \int_0^{\tau} \sqrt{n}\bigl[\bigl\{\hat S_1\bigl(\hat\bfbeta(u);u,t\bigr)
- \hat S_0\bigl(\hat\bfbeta(u);u,t\bigr)\bigr\} - \{S_1(t)-S_0(t)\}\bigr]dt \cr
&= \int_0^{\tau} \biggl[c_{01}(t) C_0(\bfbeta_0,u,t)-c_{11}(t) C_1(\bfbeta_0,u,t)
+ \mathbf{D}^T(\bfbeta_0,t) \boldsymbol{\Sigma}^{-1}(\bfbeta_0,u) {1 \over \sqrt{n}} \mathbf{U}(\bfbeta_0,u) \cr
&\qquad + \sqrt{n}\{\tilde S_1(t)-\tilde S_0(t)\} - \sqrt{n}\{S_1(t)-S_0(t)\} \biggr]dt + o_p(1).
}
The continuous mapping theorem, together with the fact that
$\sqrt{n} \bigl\{\hat S_1\bigl(\hat\bfbeta(u),u,t\bigr)- \hat S_0\bigl(\hat\bfbeta(u),u,t\bigr)\bigr\}
-\sqrt{n}\{S_1(t)-S_0(t)\} \Darrow \zeta_6(u,t)$, implies that
$$\sqrt{n} \{\hat\Delta(u,\tau)-\Delta(u,\tau)\} \Darrow \int_0^{\tau} \zeta_6(u,t)dt.$$
Therefore,
$\bigl\{\sqrt{n} \{\hat\Delta(u,\tau)-\Delta(u,\tau)\},~ u_* \leq u \leq u^* \bigr\}$
converges in distribution to a Gaussian process $\eta$
with mean 0, where $\eta(u,\tau) = \int_0^{\tau} \zeta_6(u,t)dt$.
Using the covariance function of $\zeta_6$ in \ref{eq:COVAR6},
the covariance function of $\eta$ is calculated as, for $\tau_0 \leq u_1, u_2 \leq \tau$,
\Eq{align*}
{
& K_{\eta}(u_1,u_2,\tau) = \Cov\{\eta(u_1,\tau), \eta_2(u,\tau)\} = E\{\eta(u_1,\tau)\eta(u_2,\tau)\} \cr
&\quad = E\biggl\{\int_0^{\tau} \zeta_6(u_1,s)ds \int_0^{\tau} \zeta_6(u_2,t)dt\biggr\}
= E\biggl\{\int_0^{\tau}\int_0^{\tau} \zeta_6(u_1,s)\zeta_6(u_2,t)dsdt\biggr\} \cr
&\quad= \int_0^{\tau}\int_0^{\tau} E\{\zeta_6(u_1,s)\zeta_6(u_2,t)\}dsdt
= \int_0^{\tau}\int_0^{\tau} K_{\zeta_6}\bigl((u_1,s),(u_2,t)\bigr)dsdt \cr
&\quad= \int_0^{\tau}\int_0^{\tau} [K_{\zeta_5}\bigl((u_1,s),(u_2,t)\bigr)
+ \Cov\{S_1(s|\bfZ)-S_0(s|\bfZ),~ S_1(t|\bfZ)-S_0(t|\bfZ)\}] dsdt \cr
&\quad= \int_0^{\tau}\int_0^{\tau} K_{\zeta_5}\bigl((u_1,s),(u_2,t)\bigr)dsdt \cr
&\qquad + \int_0^{\tau}\int_0^{\tau} \Cov\{S_1(s|\bfZ)-S_0(s|\bfZ),~ S_1(t|\bfZ)-S_0(t|\bfZ)\}dsdt \cr
&\quad= K_{\xi}(u_1,u_2,\tau) + K(\tau) = K_{\xi}(u_1 \vee u_2, u_1 \vee u_2, \tau) + K(\tau) \cr
&\quad = V_{\xi}^2(u_1 \vee u_2, \tau) + K(\tau),
}
where
\Eq{align*}
{
& K(\tau) = \int_0^{\tau}\int_0^{\tau} \Cov\{S_1(s|\bfZ)-S_0(s|\bfZ),~ S_1(t|\bfZ)-S_0(t|\bfZ)\}dsdt \cr
&\quad= \int_0^{\tau}\int_0^{\tau} \biggl(E[\{S_1(s|\bfZ)-S_0(s|\bfZ)\} \{S_1(t|\bfZ)-S_0(t|\bfZ)\}] \cr
&\qquad - \{S_1(s)-S_0(s\} \{S_1(t)-S_0(t)\}\biggr)dsdt \cr
&\quad= E \int_0^{\tau}\{S_1(s|\bfZ)-S_0(s|\bfZ)\}ds  \int_0^{\tau} \{S_1(t|\bfZ)-S_0(t|\bfZ)\}dt \cr
&\qquad- \int_0^{\tau} \{S_1(s)-S_0(s\}ds \int_0^{\tau}\{S_1(t)-S_0(t)\}dt \cr
&\quad= E \biggl\{\int_0^{\tau} \{S_1(t|\bfZ)-S_0(t|\bfZ)\}dt\biggr\}^2 - \{\mu_1(\tau)-\mu_0(\tau)\}^2 \cr
&\quad= \Var\biggl\{\int_0^{\tau} \{S_1(t|\bfZ)-S_0(t|\bfZ)\}dt\biggr\}
= \Var\{\mu_1(\tau|\bfZ)-\mu_0(\tau|\bfZ)\}.
}

\bibliographystyle{apalike}  
\bibliography{refs}  

\end{document}

%% file: macro.tex
\def\bt{\begin{thm}}
\def\et{\end{thm}}

\def\be{\begin{eqnarray}}
\def\ee{\end{eqnarray}}
\def\bes{\begin{eqnarray*}}
\def\ees{\end{eqnarray*}}
\def\ba{\begin{aligned}}
\def\ea{\end{aligned}}

\def\bfbeta{{\boldsymbol\beta}}

\def\parrow{\buildrel{ p}\over\longrightarrow}
\def\Darrow{\buildrel {\cal D}\over\longrightarrow}
\def\darrow{\buildrel d \over\longrightarrow}

\def\bfZ{{\bf Z}}
\def\bfz{{\bf z}}
\def\bfY{{\bf Y}}
\def\bfy{{\bf y}}
\def\Var{{\rm Var}}
\def\Cov{{\rm Cov}}

\def\cI{{\cal I}}

\def\bd{\begin{description}}
\def\ed{\end{description}}

\newcommand{\Eq}[2]{\begin{#1}#2\end{#1}}